\documentclass[lettersize,journal]{IEEEtran}
\usepackage{amsmath,amsfonts}
\usepackage{algorithmic}
\usepackage{algorithm}
\usepackage{array}
\usepackage[caption=false,font=normalsize,labelfont=sf,textfont=sf]{subfig}
\usepackage{textcomp}
\usepackage{stfloats}
\usepackage{url}
\usepackage{hyperref}
\usepackage{verbatim}
\usepackage{graphicx}
\usepackage{makecell}
\usepackage{cite}
\usepackage{color}
\usepackage{booktabs}
\newcolumntype{I}{!{\vrule width 1.5pt}}
\newlength\savedwidth
\newcommand\whline{\noalign{\global\savedwidth\arrayrulewidth
                           \global\arrayrulewidth 2pt}%
                  \hline
                  \noalign{\global\arrayrulewidth\savedwidth}}
\newlength\savewidth
\newcommand\shline{\noalign{\global\savewidth\arrayrulewidth
                           \global\arrayrulewidth 1.5pt}%
                  \hline
                  \noalign{\global\arrayrulewidth\savewidth}}

\begin{document}

%\title{A Survey on Multimodal Recommender Systems: Technological Taxonomy and Future Directions}
\title{A Survey on Multimodal Recommender Systems: Recent Advances and Future Directions}

% \author{Jinfeng Xu,~\IEEEmembership{Student Member,~IEEE,}
        % <-this % stops a space
\author{Jinfeng Xu,
        Zheyu Chen,
        Shuo Yang,
        Jinze Li,
        Wei Wang,\\
        Xiping Hu,
        Steven Hoi,~\IEEEmembership{Fellow,~IEEE} 
        and Edith Ngai,~\IEEEmembership{Senior Member,~IEEE}% <-this % stops a space
\thanks{Jinfeng Xu, Jinze Li, Shuo Yang, and Edith C. H. Ngai* are with the Department of Electrical and Electronic Engineering, The University of Hong Kong, Hong Kong, China (e-mail: {jinfeng, lijinze-hku, shuo.yang,}@connect.hku.hk, chngai@eee.hku.hk).}% <-this % stops a space
\thanks{Zheyu Chen is with the Department of Electrical and Electronic Engineering, The Hong Kong Polytechnic University, Hong Kong, China (e-mail: zheyu.chen@connect.polyu.hk).}% <-this % stops a space
\thanks{Wei Wang and Xiping Hu are with the Department of Engineering, Shenzhen MSU-BIT University, Shenzhen, China, and also with the School of Medical Technology, Beijing Institute of Technology, Beijing, China (e-mail: ehomewang@ieee.org, huxp@bit.edu.cn).}

\thanks{Steven Hoi is with the School of Computing and Information Systems, Singapore
Management University, Singapore (e-mail: chhoi@smu.edu.sg).}% <-this % stops a space
\thanks{*The corresponding author.}% <-this % stops a space
\thanks{This paper was produced by the IEEE Publication Technology Group. They are in Piscataway, NJ.}% <-this % stops a space
\thanks{Manuscript received XXXX; revised XXXX.}}

% % The paper headers
\markboth{IEEE TRANSACTION ON MULTIMEDIA,~Vol.~X, No.~X, X~2025}%
{Shell \MakeLowercase{\textit{et al.}}: IEEE TRANSACTION ON MULTIMEDIA}

% \IEEEpubid{0000--0000/00\$00.00~\copyright~2021 IEEE}
% Remember, if you use this you must call \IEEEpubidadjcol in the second
% column for its text to clear the IEEEpubid mark.

\maketitle

\begin{abstract}
Acquiring valuable data from the rapidly expanding information on the internet has become a significant concern, and recommender systems have emerged as a widely used and effective tool for helping users discover items of interest. The essence of recommender systems lies in their ability to predict users' ratings or preferences for various items and subsequently recommend the most relevant ones based on historical interaction data and publicly available information. With the advent of diverse multimedia services, including text, images, video, and audio, humans can perceive the world through multiple modalities. Consequently, a recommender system capable of understanding and interpreting different modal data can more effectively refer to individual preferences. Multimodal Recommender Systems (MRS) not only capture implicit interaction information across multiple modalities but also have the potential to uncover hidden relationships between these modalities. The primary objective of this survey is to comprehensively review recent research advancements in MRS and to analyze the models from a technical perspective. Specifically, we aim to summarize the general process and main challenges of MRS from a technical perspective. We then introduce the existing MRS models by categorizing them into four key areas: Feature Extraction, Encoder, Multimodal Fusion, and Loss Function. Finally, we further discuss potential future directions for developing and enhancing MRS. This survey serves as a comprehensive guide for researchers and practitioners in MRS field, providing insights into the current state of MRS technology and identifying areas for future research. We hope to contribute to developing a more sophisticated and effective multimodal recommender system. To access more details of this paper, we open source a repository: \href{https://github.com/Jinfeng-Xu/Awesome-Multimodal-Recommender-Systems}{https://github.com/Jinfeng-Xu/Awesome-Multimodal-Recommender-Systems}.
\end{abstract}


\section{Introduction}
\IEEEPARstart{T}{he} rapid expansion of the Internet has resulted in an overwhelming abundance of information, making it increasingly challenging for users to identify what is useful and relevant. This phenomenon, referred to as information overload, arises due to the near impossibility of controlling the generation and dissemination of information in the digital age. Consequently, there is an urgent need for robust filtering mechanisms that prioritize pertinent content to facilitate efficient communication and decision-making processes. Recommender systems, which personalize content filters according to specific requirements across various domains, have demonstrated their efficacy in mitigating the adverse effects of information overload. These systems have proven particularly successful in commercial applications such as e-commerce, advertising, and social media, where personalization is crucial to user engagement and satisfaction \cite{deldjoo2021content,deldjoo2022review,xu2024fourierkangcf,xu2024aligngroup}.

The primary function of recommender systems is to predict users' ratings or preferences for various items and recommend the most likely and relevant items based on historical interaction data and publicly available information. However, traditional ID-based recommendation methods, which operate on the principle that users tend to select items akin to those they have previously liked, often strongly depend on enough user-item interactions. Despite their successes, recommender systems face two significant challenges: data sparsity and the cold start problem. Data sparsity arises from the natural sparsity of interaction data between users and products, making it difficult to accurately predict users' preferences. This sparsity can lead to unreliable recommendations, especially in systems with large item catalogs but relatively few user interactions. The cold start problem occurs because traditional recommender system models rely heavily on ID embeddings, which struggle to make satisfactory predictions for new users or products with little to no historical interaction data. This challenge is particularly pronounced in dynamic environments where new items and users are continuously introduced.

To alleviate these issues, multimodal information is increasingly being integrated into recommendation systems. MRS leverages auxiliary multimodal information, such as text, images, videos, and audio, to complement the historical interactions between users and items. This approach enhances recommendation performance by providing a richer and more comprehensive understanding of user preferences. The essential goal of recommender systems is to cater to people's preferences, and since human perception of the world is inherently multimodal, incorporating diverse modal information can capture preferences at a finer level of granularity. This leads to more accurate and personalized recommendations, thereby improving user satisfaction and engagement.

Research in multimodal recommendation is rapidly growing and evolving. To assist researchers in quickly understanding MRS and to support community development, a comprehensive review from a technical perspective is urgently needed. Existing work \cite{liu2023multimodal} attempts to categorize MRS from a technical standpoint; however, the rapid advancement of the field has rendered some of its categorizations outdated. Therefore, we aim to collect recent work and propose a more up-to-date categorization framework to help researchers grasp the latest progress in the MRS community. This review will provide a thorough overview of current MRS technologies, highlight emerging trends, and identify potential future directions for research and development in this dynamic field. By systematically examining the state-of-the-art (SOTA) works in MRS, we hope to contribute to the ongoing efforts to enhance the capabilities and applications of recommender systems in a multimodal digital world.

\begin{table*}[!ht]
    \centering
    \caption{Related recommender system survey}
    \begin{tabular}{|m{3cm}<{\centering}|m{7cm}<{\centering}|m{7cm}<{\centering}|}
    \hline
        Surveys & Key Contributions & Differences from Our Survey \\ \hline
        Zhang et al.\cite{zhang2019deep} & provide a panorama for advances in deep learning based recommender systems and provide a survey of future direction and challenges. & This work comprehensively demonstrates the advanced development in deep learning based recommender systems, which include MRS but lacks the fine-grained introduction to existing state-of-the-art technologies. \\ \hline
        Guo et al. \cite{guo2020survey} & provide a fine-grained survey for the existing approaches utilizing the KG to improve the recommendation result and introduce some datasets used in different scenarios. & Our survey focuses on the taxonomy of the process for MRS and state-of-the-art technologies of the multimodal recommender system; the KG only discussed as a part of the techniques in our work. \\ \hline
        Deldjoo et al. \cite{deldjoo2020recommender} & provide a comprehensive and coarse-grained survey and a coarse-grained categorization by the modalities. & The categorization in this work is coarse-grained and unreasonable to some extent, while our work provides a fine-grained categorization for techniques. \\ \hline
        Jannach et al. \cite{jannach2021survey} & explore the field of CRS and provide a taxonomic survey of existing techniques. & This work discusses the recent approaches in the CRS field but lacks a combination with multimodal information. \\ \hline
        Deldjoo et al. \cite{deldjoo2021content} & discuss the state-of-the-art approaches to content-driven MRS and provide a survey of challenges and historical evolution. & This work focuses on the content-driven MRS rather than covering all perspectives of MRS. \\ \hline
        Wu et al. \cite{wu2022graph} & provide a comprehensive survey for utilizing GNN techniques in the RS field and list several limitations and future directions. & Our survey focuses on the MRS field and provides a more fine-grained classification of GNN techniques in the MRS field. \\ \hline
        Deldjoo et al. \cite{deldjoo2022review} & provide a comprehensive survey of RS in the fashion field according to the task in the market, and provide some vital evaluation goals in the fashion field. & This work focuses on the RS in the fashion field but lacks a general and fine-grained survey for RS. \\ \hline
        Meng et al. \cite{meng2023survey} & provide a comprehensive analysis for personalized news recommendations via technologies and list several limitations and future directions. & Our survey focuses on the taxonomy of the process for general MRS and state-of-the-art technologies, which are also effective in the news field. \\ \hline
        Zhou et al. \cite{zhou2023comprehensive} & summarize the main methods used in MRS and provide a common framework for commonly used MRS models. & This work comprehensively demonstrates the previous approaches in MRS but lacks a fine-grained process of MRS. \\ \hline 
        Liu et al. \cite{liu2023multimodal} & summarize the main methods used in MRS and provide a common framework for commonly used MRS models. & This work delineates the MRS approach from a process perspective, rather than a technology development perspective, and does not allow the reader to fully understand the direction of research in the field. \\ \hline 
    \end{tabular}
    \label{tab: Related recommender system survey}
\end{table*}

\subsection{Search Strategy for Relevant Papers}
We conducted a comprehensive survey on Multimodal Recommendation Systems (MRS) by systematically retrieving and analyzing articles from leading conferences and journals in the field. The conferences and journals included, but were not limited to, MM, KDD, WWW, SIGIR, AAAI, ICLR, IJCAI, CIKM, WSDM, TMM, TKDE, TPAMI, and INFFUS. This rigorous selection process ensured that our survey covered the most influential and cutting-edge research in MRS.

Our search approach was methodically divided into three distinct stages:

\begin{itemize}
    \item Collection of High-Quality Articles: In the initial stage, we gathered articles from the aforementioned top conferences and journals. This selection was based on the reputation and impact factor of the sources, ensuring that only high-quality and peer-reviewed research was included in our survey.
    \item Filtering and Post-Processing: Following the collection phase, we meticulously filtered and post-processed the articles. This step involved removing duplicates, assessing the relevance of each article to the topic of MRS, and ensuring that only the most pertinent studies were retained. This rigorous filtering process was crucial for maintaining the focus and quality of our survey.
    \item Technical Analysis and Synthesis: In the final stage, we conducted a detailed analysis of the techniques employed in each article. This involved examining the methodologies, models, and algorithms used, as well as the motivations behind these approaches. We also reviewed related works cited within each article to provide a comprehensive understanding of the evolution and current trends in MRS. By synthesizing this information, we were able to summarize the key techniques and motivations driving the field.
\end{itemize}
Through this systematic approach, our survey offers a thorough and insightful overview of the SOTA works in Multimodal Recommendation Systems. It highlights the significant advancements, emerging trends, and potential future directions in the field, providing valuable guidance for researchers and practitioners alike.

\subsection{Compared with Related Surveys}
Several surveys have been published on recommender systems, addressing either general aspects or specific facets of these systems. However, none provide a comprehensive and reasonable taxonomy of the processes and detailed technologies utilized in recent SOTA MRS works, which is an emerging and crucial requirement in this field. The objective of MRS is to enhance the capability of extracting deeper and more accurate interactions between users and items by incorporating multimodal information into recommender systems. This paper discusses the main contributions and limitations of existing related surveys and highlights the unique contributions of our work, as summarized in Table \ref{tab: Related recommender system survey}.

Zhang et al. \cite{zhang2019deep} offer a panoramic view of advances in deep learning-based recommender systems, surveying future directions and challenges, including joint representation learning, explainability, deeper models, and machine reasoning. However, their work lacks a fine-grained introduction to existing SOTA technologies. Deldjoo et al. \cite{deldjoo2020recommender} provide a comprehensive survey and a coarse-grained categorization by modalities, including common features such as audio, visual, and textual, as well as special features like motion, metadata, and semantic orientation. Nonetheless, this categorization is somewhat coarse-grained and lacks precision.

Jannach et al. \cite{jannach2021survey} explore the field of conversational recommender systems (CRS) and offer a taxonomic survey of existing techniques, but their work does not integrate multimodal information. Deldjoo et al. \cite{deldjoo2021content} discuss SOTA approaches to content-driven MRS, surveying challenges and historical evolution, including increasing recommendation diversity and novelty, providing transparency and explanations, achieving context-awareness, improving scalability and efficiency, and alleviating the cold start problem. However, their focus is primarily on content-driven MRS, rather than covering the general MRS landscape.

Previous works \cite{guo2020survey,wu2022graph} focus on graph structure in recommendation systems. Guo et al. \cite{guo2020survey} provide a fine-grained survey of approaches utilizing knowledge graphs (KG) to enhance recommendation results, categorizing methods into embedding-based, path-based, and unified approaches. Wu et al. \cite{wu2022graph} offer a comprehensive survey of graph neural network (GNN) techniques in recommender systems, identifying several limitations and future directions, including diverse and uncertain representation, scalability, dynamics, receptive fields, self-supervised learning, robustness, privacy-preserving methods, and fairness.

Deldjoo et al. \cite{deldjoo2022review} provide a comprehensive survey of recommender systems in the fashion domain, categorizing tasks in the market and outlining vital evaluation goals specific to fashion. Meng et al. \cite{meng2023survey} present a thorough analysis of personalized news recommendations, discussing technologies and listing several limitations and future directions, including privacy protection, fake news mitigation, and de-biasing.

Zhou et al. \cite{zhou2023comprehensive} summarize the main methods employed in MRS and propose a common framework for commonly used MRS models. While their work offers a comprehensive overview of previous approaches in MRS, the pipeline proposed for MRS requires more detailed elaboration. More recently, Liu et al. \cite{liu2023multimodal} also summarize the main methods used in MRS and provide a common framework for MRS models. However, this work delineates MRS from a process perspective rather than focusing on technological developments, limiting the reader’s ability to fully understand the research directions in the field.

In conclusion, our work aims to fill these gaps by providing a more detailed and up-to-date taxonomy of MRS processes and technologies, thereby advancing the understanding and development of this rapidly evolving domain.

Our survey focuses on a refined categorization of MRS from a technological perspective to provide researchers with insight into the technological development of MRS. Finally, we discuss potential future directions for developing and improving multimodal recommendations.

\subsection{The Outline of the Survey}

The structure of our survey is organized according to the following:
\begin{itemize}
    \item \textbf{Section I: Introduction}
    \\ We briefly outline the historical development of RS and underscore the significance of leveraging multimodal information to enhance recommendations. Subsequently, we detail the search strategy employed to ensure the quality of our work. Additionally, we offer a comparative analysis with previous surveys. Finally, we present the structure of this survey and highlight the main contributions of our research.
    \item \textbf{Section II: Technological Taxonomy}
    \\We present a technological taxonomy for MRS summarize all recent SOTA works, and then overview the four key technologies respectively.
    \item \textbf{Section III: Feature Extraction}
    \\We have comprehensively reviewed recent MRS works in feature extraction techniques for visual and textual modalities, identifying prevailing trends. This analysis has culminated in a curated standard setting, sparing researchers the complexity of navigating diverse methods.
    \item \textbf{Section IV: Encoder}
    \\We categorize the MRS encoder types from a technical perspective and summarize all recent SOTA works. This comprehensive overview provides insights into potential research gaps and future directions, facilitating ongoing development and refinement in MRS technologies.
    \item \textbf{Section V: Multimodal Fusion}
    \\We provide a comprehensive categorization for multimodal fusion in MRS, from both timing and strategy perspectives. We also summarize all recent works from these perspectives and reveal the interplays between these two perspectives. This comprehensive categorization is intended to help researchers select appropriate timing and strategies for modal fusion.
    \item \textbf{Section VI: Loss Function}
    \\We provide a detailed introduction to loss functions utilized in MRS, including both supervised and self-supervised learning frameworks. Specifically, we introduce and analyze several widely used loss functions, detailing their applications and effectiveness across different learning scenarios within MRS. This exploration aims to assist researchers in selecting and implementing the most suitable loss functions.
    % \item \textbf{Section VII: Literature Review}
    % \\We conduct a detailed literature review of SOTA MRS works from a technical perspective. We hope to help researchers and practitioners looking to understand current trends and future directions in MRS.
    \item \textbf{Section VII: Future Direction}
    \\We delve into the potential future directions of the MRS field. We aim to stimulate and encourage further research, development, and innovation in this rapidly evolving area of research.
    \item \textbf{Section VIII: Conclusion}
    \\We briefly summarize the contents and contributions of this survey. 
\end{itemize}

\begin{figure}[!t]
\centering
\includegraphics[width=3.5in]{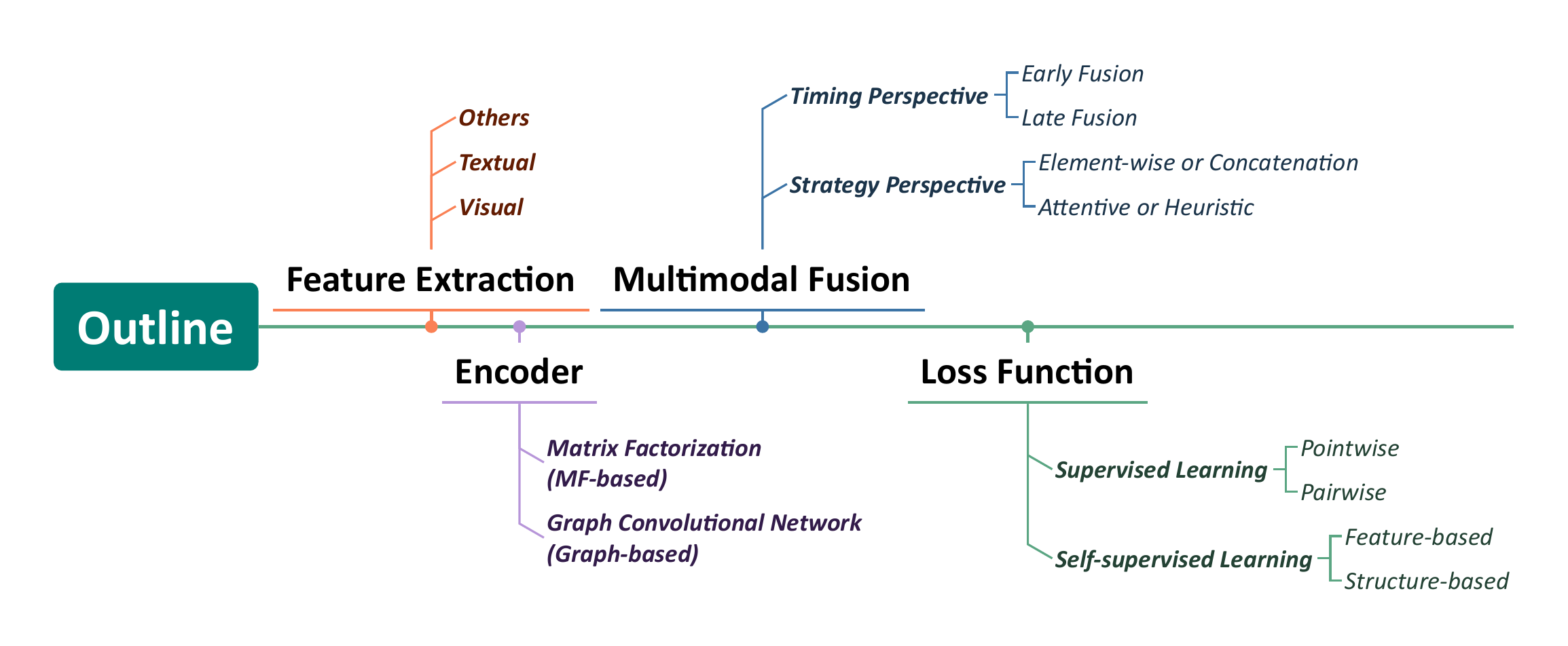}
\caption{The outline structure of our survey.}
\label{structure}
\end{figure}

Figure~\ref{structure} illustrates a detailed outline structure of our survey (Including Section III - Section VI).

The main contributions of our survey are as follows:
\begin{itemize}
    \item We provide a comprehensive review of MRS, summarize the huge number of SOTA works, and structure a general process of MRS to present how multimodal information is utilized in the RS field.
    \item We categorize and analyze the techniques and motivations for each main step of this general process of MRS, which can extremely guide the researcher to conduct further research.
    \item We introduce the commonly available datasets for MRS and provide a detailed characterization of them. And organized the datasets used in SOTA works to help researchers choose the suitable dataset.
    \item We discuss the existing challenges for MRS based on previous works and list some future directions, which is worthy of in-depth research. 
\end{itemize}

\begin{table}[!ht]
    \centering
    \caption{List of abbreviations used throughout this paper}
    \begin{tabular}{cc|l}
    \hline
         \makecell[c]{\textbf{Abbreviation}} & ~ & \makecell[c]{\textbf{Term}}  \\ \hline
         \makecell[l]{RS} & ~ & Recommender Systems  \\ \hline
         \makecell[l]{MRS} & ~ & Multimodal Recommender Systems  \\ \hline
         \makecell[l]{CRS} & ~ & Conversational Recommender Systems  \\ \hline 
         \makecell[l]{CF} & ~ & Collaborative Filtering  \\ \hline
         \makecell[l]{MF} & ~ & Matrix Factorization  \\ \hline
         \makecell[l]{GCN} & ~ & Graph Convolution Network  \\ \hline
         \makecell[l]{GNN} & ~ & Graph Neural Network  \\ \hline
         \makecell[l]{KG} & ~ & Knowledge Graphs  \\ \hline
         \makecell[l]{SOTA} & ~ & State-of-the-art  \\ \hline
         % \makecell[l]{DRL} & ~ & Disentangled Representation Learning  \\ \hline
         \makecell[l]{CL} & ~ & Contrastive Learning  \\ \hline
         \makecell[l]{SSL} & ~ & Self-supervised Learning  \\ \hline
         \hline
    \end{tabular}
    \label{tab: List of abbreviations used throughout this paper}
\end{table}

Table~\ref{tab: List of abbreviations used throughout this paper} lists the abbreviations used throughout this paper.
\section{Technological Taxonomy}

\begin{figure*}[!t]
\centering
\includegraphics[width=1\linewidth]{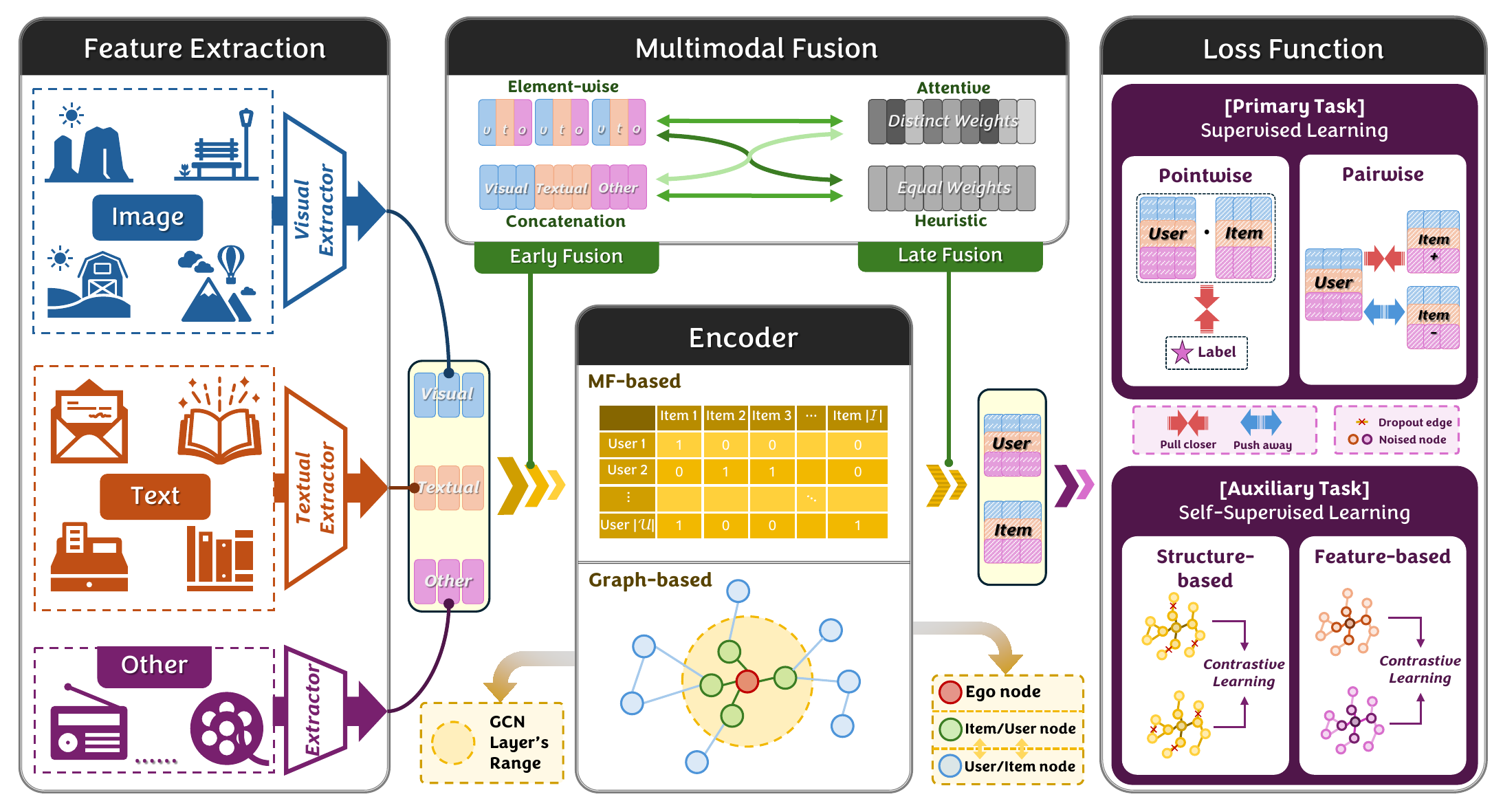}
\caption{Technologies in MRS from process pipeline perspectives.}
\label{MRS}
\end{figure*}
Based on the current MRS work at SOTA to summarize and organize, we classify the technologies in MRS into four parts as shown in Figure~\ref{MRS}. Specifically, there are four parts: \textbf{Feature Extraction}, \textbf{Encoder}, \textbf{Multimodal Fusion}, and \textbf{Loss Function}. We briefly overview these parts and discuss them in detail in the subsequent sections. 

\subsection{Feature Extraction} 
Different application scenarios encompass varying types of modality information, leading to diverse datasets with distinct multimodal features. Most datasets, however, include at least three primary modalities: interaction, visual, and textual. For instance, large platforms such as Amazon, Netflix, and TikTok provide datasets rich in image and textual information, thus covering both visual and textual modalities. Specifically, TikTok datasets often include additional modalities, such as audio and video \cite{wei2019mmgcn,wang2021dualgnn,zhang2021mining}. Furthermore, datasets from specialized domains sometimes have rare modalities. For example, datasets in popular areas like fashion and healthcare often include a variety of specialized modalities.

Feature extraction is a critical process aimed at the representation of low-dimensional, interpretable channel features using embedding techniques. Different pre-extraction methods are employed for distinct modalities. For the visual modality, models such as ResNet \cite{he2016deep} and ViT \cite{dosovitskiy2020image} are utilized to extract features. In the case of the textual modality, models like BERT \cite{devlin2018bert} and Sentence-Transformer \cite{reimers2019sentence} are used to derive features. Audio features are typically extracted using models such as LSTM \cite{hochreiter1997long} and GRU \cite{cho2014properties}.

A detailed introduction to feature extraction is provided in Section~\ref{sec: Feature Extraction}, where we delve into the specifics of each modality and the corresponding extraction techniques. 

\subsection{Encoder}
The encoder utilizes features extracted from multimodal information and historical interaction data to infer user preference representations, which are subsequently used in predicting user-item interactions for making recommendations. Similar to traditional recommender systems, encoders for multimodal recommendation can be broadly classified into \textbf{Matrix Factorization (MF \cite{mnih2007probabilistic})-based} and \textbf{Graph Convolutional Network (Graph \cite{wu2019simplifying})-based} approaches. The MF-based approach is known for its simplicity and effectiveness, whereas the Graph-based approach leverages the bipartite graph inherent in user-item interactions to learn higher-order neighbor features.

With the rapid advancement of MRS, more sophisticated encoders have been proposed and utilized to fully exploit the rich multimodal information, thereby enhancing recommendation performance. These advanced encoders enable the integration of diverse multimodal data, leading to more accurate and personalized recommendations.

In Section~\ref{sec: Feature Extraction}, we will provide a detailed introduction to the development and motivations behind both types of encoders. This includes an exploration of how MF-based methods efficiently capture user-item interactions and how Graph-based methods extend this capability by incorporating complex graph structures. By examining these methodologies, we aim to elucidate the strengths and limitations of each approach, as well as their contributions to the MRS community.

\subsection{Multimodal Fusion}
One of the key research focuses in MRS is the Multimodal Fusion. Recent studies have demonstrated that the timing of modality fusion can significantly impact the effectiveness of the recommendations. Multimodal fusion involves integrating information from different modalities at various stages, and this timing can be crucial for achieving optimal performance.

\textbf{Early Fusion}: Early fusion involves combining different modality features before they are processed by the encoder. This approach can effectively uncover hidden relationships between modalities, as the integrated multimodal features allow the encoder to learn richer and higher-quality representations. Early fusion can capture the intricate interactions between different types of data, such as text, images, and audio, leading to a more holistic understanding of user preferences. Techniques for early fusion often include concatenation, attention mechanisms, and neural network-based integration methods, which aim to create a unified representation of multimodal data.

\textbf{Late Fusion}: Late fusion combines the scores or predictions from each modality after the individual modality-specific encoders have processed them. This approach focuses on leveraging the strengths of each modality-specific model and then combining their outputs to make the final recommendation. Late fusion can be particularly effective in scenarios where certain modalities are more informative or reliable than others. By deferring the fusion process until after the prediction phase, late fusion allows for more targeted and refined extraction of specific modality information, enhancing the overall recommendation accuracy.

In Section~\ref{sec: Multimodal Fusion}, we will provide a detailed classification of existing work based on the timing of fusion, categorizing them into early fusion and late fusion approaches. This classification will offer a comprehensive understanding of how different fusion strategies impact the performance of MRS systems. We will explore various methodologies and techniques employed in both early and late fusion, analyzing their advantages, limitations, and application scenarios.

\subsection{Loss Function}
MRS leverages loss functions that can be broadly divided into two components: primary tasks and auxiliary tasks. The primary tasks are supervised learning, which typically involves clearly defined labels to guide the model's learning process. These tasks ensure that the model learns to make accurate predictions based on labeled data. The auxiliary tasks are self-supervised learning (SSL) \cite{liu2021self}. SSL generates supervision signals from the inherent structure or patterns within the data itself, rather than relying solely on externally labeled data. This approach allows recommender systems to utilize unlabeled data effectively, extracting meaningful representations and making accurate predictions even in data sparsity scenarios.

Supervised Learning can be further subdivided into Pointwise Loss and Pairwise Loss:

\textbf{Pointwise Loss}: This loss is calculated by comparing the predicted score for each individual item with its actual label. Common pointwise loss functions include Mean Squared Error (MSE) \cite{marmolin1986subjective} and Cross-Entropy Loss (CE) \cite{lecun2015deep}, which are used to directly assess the accuracy of individual predictions.

\textbf{Pairwise Loss}: This loss focuses on the relative ranking of items. It evaluates the model's ability to correctly order each pair of items based on user preferences. Common pairwise loss functions include Bayesian Personalized Ranking (BPR) \cite{rendle2012bpr} and Hinge Loss \cite{cortes1995support}, which aim to optimize the rank order of items rather than their absolute scores.

Self-supervised Learning can be categorized into feature-based and structure-based methods:

\textbf{Feature-based SSL}: This method involves creating auxiliary tasks that predict or reconstruct certain features of the data. For example, a model might be trained to predict missing features of an item or user based on the available data, thereby learning more robust representations.

\textbf{Structure-based SSL}: This approach leverages the structural properties of the data, such as the relationships and interactions between users and items. Graph-based methods, for instance, might use node similarity or subgraph patterns to generate supervision signals, enhancing the model's ability to capture complex dependencies and interactions.

In Section~\ref{sec: Loss Function}, we will provide a detailed introduction to these loss functions. We will explore the motivations behind each type of loss, their implementation details, and their impact on the performance of multimodal recommender systems. By examining both supervised and self-supervised learning strategies, we aim to offer a comprehensive understanding of how different loss functions contribute to the effectiveness of multimodal recommendation.

\section{Feature Extraction}
\label{sec: Feature Extraction}

\begin{table*}[!ht]
    \centering
    \caption{Summary of MRS feature extraction for both visual and textual modalities.}
    \begin{tabular}{IcccIc|c|c|c|cIc|c|c|c|c|c|c|c|c|c|cI}
    \whline
    \noalign{\smallskip}
        \multicolumn{3}{c}{Modality} & \multicolumn{5}{c}{Visual} & \multicolumn{11}{c}{Textual}\\
        \cmidrule(rl){1-3}   \cmidrule(rl){4-8}  \cmidrule(rl){9-19}
        
        \textbf{Method} & Year & Publication & \rotatebox{90}{Provided} & \rotatebox{90}{VGG} & \rotatebox{90}{Inception} & \rotatebox{90}{Caffe} & \rotatebox{90}{ResNet} & \rotatebox{90}{Provided} & \rotatebox{90}{TF-IDF} & \rotatebox{90}{GRU} & \rotatebox{90}{PV-DM} & \rotatebox{90}{PV-DBOW} & \rotatebox{90}{Glove} & \rotatebox{90}{Attention} & \rotatebox{90}{BERT} & \rotatebox{90}{Word2Vec} & \rotatebox{90}{\makecell[c]{Sentence- \\ transformer}} & \rotatebox{90}{\makecell[c]{Sentence \\ 2Vec}} \\ \shline
        \textbf{(V)VBPR\cite{he2016vbpr}} & 2016 & AAAI & \checkmark & ~ & ~ & ~ & ~ & ~ & ~ & ~ & ~ & ~ & ~ & ~ & ~ & ~ & ~ &\\ \hline
        
        \textbf{(V)VMCF\cite{park2017also}} & 2017 & WWW & \checkmark & ~ & ~ & ~ & ~ & ~ & ~ & ~ & ~ & ~ & ~ & ~ & ~ & ~ & ~ &\\ \hline
        \textbf{(V)ACF\cite{chen2017attentive}} & 2017 & SIGIR & ~ & ~ & ~ & ~ & \checkmark & ~ & ~ & ~ & ~ & ~ & ~ & ~ & ~ & ~ & ~ &\\ \hline
        \textbf{JRL\cite{zhang2017joint}} & 2017 & CIKM & \checkmark & ~ & ~ & ~ & ~ & ~ & ~ & ~ & ~ & \checkmark & ~ & ~ & ~ & ~ & ~ &\\ \hline
        \textbf{(V)DVBPR\cite{kang2017visually}} & 2017 & ICDM & ~ & ~ & ~ & ~ & \checkmark & ~ & ~ & ~ & ~ & ~ & ~ & ~ & ~ & ~ & ~ &\\ \hline
        
        \textbf{GraphCAR\cite{xu2018graphcar}} & 2018 & SIGIR & \checkmark & ~ & ~ & ~ & ~ & \checkmark & ~ & ~ & ~ & ~ & ~ & ~ & ~ & ~ & ~ &\\ \hline
        
        \textbf{VECF\cite{chen2019personalized}} & 2019 & SIGIR & ~ & \checkmark & ~ & ~ & ~ & ~ & ~ & \checkmark & ~ & ~ & ~ & ~ & ~ & ~ & ~ &\\ \hline
        \textbf{UVCAN\cite{liu2019user}} & 2019 & WWW & ~ & ~ & \checkmark & ~ & ~ & ~ & ~ & ~ & ~ & ~ & ~ & ~ & ~ & ~ & ~ &\\ \hline
        \textbf{MAML\cite{liu2019user2}} & 2019 & MM & ~ & ~ & ~ & \checkmark & ~ & ~ & ~ & ~ & \checkmark & ~ & ~ & ~ & ~ & ~ & ~ & \\ \hline
        \textbf{MMGCN\cite{wei2019mmgcn}} & 2019 & MM & \checkmark & ~ & ~ & ~ & ~ & \checkmark & ~ & ~ & ~ & ~ & ~ & ~ & ~ & ~ & ~ & \\ \hline
        \textbf{(V)AMR\cite{tang2019adversarial}} & 2019 & TKDE & ~ & ~ & ~ & ~ & \checkmark & ~ & ~ & ~ & ~ & ~ & ~ & ~ & ~ & ~ & ~ & \\ \hline
        
        \textbf{MGAT\cite{tao2020mgat}} & 2020 & I\&M & \checkmark & ~ & ~ & ~ & ~ & \checkmark & ~ & ~ & ~ & ~ & ~ & ~ & ~ & ~ & ~ & \\ \hline
        \textbf{GRCN\cite{wei2020graph}} & 2020 & MM & \checkmark & ~ & ~ & ~ & ~ & \checkmark & ~ & ~ & ~ & ~ & ~ & ~ & ~ & ~ & ~ & \\ \hline
        \textbf{MKGAT\cite{sun2020multi}} & 2020 & CIKM & ~ & ~ & ~ & ~ & \checkmark & ~ & ~ & ~ & ~ & ~ & ~ & ~ & ~ & ~ & \checkmark & \\ \hline
        
        \textbf{IMRec\cite{xun2021we}} & 2021 & MM & \checkmark & ~ & ~ & ~ & ~ & ~ & ~ & ~ & ~ & ~ & ~ & \checkmark & ~ & ~ & ~ & \\ \hline
        \textbf{PMGT\cite{liu2021pre}} & 2021 & MM & ~ & ~ & \checkmark & ~ & ~ & ~ & ~ & ~ & ~ & ~ & ~ & ~ & \checkmark & ~ & ~ & \\ \hline
        \textbf{LATTICE\cite{zhang2021mining}} & 2021 & MM & \checkmark & ~ & ~ & ~ & ~ & ~ & ~ & ~ & ~ & ~ & ~ & ~ & ~ & ~ & \checkmark & \\ \hline
        \textbf{HHFAN\cite{cai2021heterogeneous}} & 2021 & TMM & \checkmark & ~ & ~ & ~ & ~ & \checkmark & ~ & ~ & ~ & ~ & ~ & ~ & ~ & ~ & ~ & \\ \hline
        \textbf{MVGAE\cite{yi2021multi}} & 2021 & TMM & ~ & ~ & ~ & ~ & \checkmark & ~ & ~ & ~ & ~ & ~ & ~ & ~ & ~ & ~ & ~ & \checkmark \\ \hline
        \textbf{DualGNN\cite{wang2021dualgnn}} & 2021 & TMM & \checkmark & ~ & ~ & ~ & ~ & \checkmark & ~ & ~ & ~ & ~ & ~ & ~ & ~ & ~ & ~ & ~ \\ \hline

        \textbf{PAMD\cite{han2022modality}} & 2022 & WWW & ~ & \checkmark & ~ & ~ & ~ & ~ & ~ & ~ & ~ & ~ & \checkmark & ~ & ~ & ~ & ~ & ~ \\ \hline
        \textbf{MMGCL\cite{yi2022multi}} & 2022 & SIGIR & \checkmark & ~ & ~ & ~ & ~ & \checkmark & ~ & ~ & ~ & ~ & ~ & ~ & ~ & ~ & ~ & ~ \\ \hline
        \textbf{(T)ADDVAE\cite{tran2022aligning}} & 2022 & KDD & ~ & ~ & ~ & ~ & ~ & ~ & \checkmark & ~ & ~ & ~ & ~ & ~ & ~ & ~ & ~ & ~ \\ \hline
        \textbf{EliMRec\cite{liu2022elimrec}} & 2022 & MM & ~ & ~ & ~ & ~ & \checkmark & ~ & ~ & ~ & ~ & ~ & ~ & ~ & ~ & ~ & ~ & \checkmark \\ \hline 
        \textbf{EgoGCN\cite{chen2022breaking}} & 2022 & MM & \checkmark & ~ & ~ & ~ & ~ & ~ & ~ & ~ & ~ & ~ & ~ & ~ & ~ & ~ & ~ & \checkmark \\ \hline
        \textbf{InvRL\cite{du2022invariant}} & 2022 & MM & ~ & ~ & ~ & ~ & \checkmark & ~ & ~ & ~ & ~ & ~ & ~ & ~ & ~ & ~ & ~ & \checkmark \\ \hline
        \textbf{A2BM2GL\cite{cai2022adaptive}} & 2022 & MM & \checkmark & ~ & ~ & ~ & ~ & \checkmark & ~ & ~ & ~ & ~ & ~ & ~ & ~ & ~ & ~ & ~ \\ \hline
        \textbf{HCGCN\cite{mu2022learning}} & 2022 & MM & \checkmark & ~ & ~ & ~ & ~ & ~ & ~ & ~ & ~ & ~ & ~ & ~ & ~ & ~ & \checkmark & ~ \\ \hline
        \textbf{SLMRec\cite{tao2022self}} & 2022 & TMM & \checkmark & ~ & ~ & ~ & ~ & \checkmark & ~ & ~ & ~ & ~ & ~ & ~ & ~ & ~ & ~ & ~ \\ \hline
        \textbf{DMRL\cite{liu2022disentangled}} & 2022 & TMM & ~ & ~ & ~ & ~ & \checkmark & ~ & ~ & ~ & ~ & ~ & ~ & ~ & ~ & \checkmark & ~ & ~ \\ \hline
         % \textbf{LUDP\cite{lei2023learning}} & 2022 & TOMCCAP & \checkmark & ~ & ~ & ~ & ~ & ~ & \checkmark & ~ & ~ & ~ & ~ & ~ & ~ & ~ & ~ & ~ & ~ \\ \hline

        \textbf{BM3\cite{zhou2023bootstrap}} & 2023 & WWW & \checkmark & ~ & ~ & ~ & ~ & ~ & ~ & ~ & ~ & ~ & ~ & ~ & ~ & ~ & \checkmark & ~ \\ \hline
        \textbf{MMSSL\cite{wei2023multi}} & 2023 & WWW & \checkmark & ~ & ~ & ~ & ~ & ~ & ~ & ~ & ~ & ~ & ~ & ~ & \checkmark & ~ & ~ & ~ \\ \hline
        \textbf{BCCL\cite{yang2023modal}} & 2023 & MM & \checkmark & ~ & ~ & ~ & ~ & ~ & ~ & ~ & ~ & ~ & ~ & ~ & \checkmark & ~ & ~ & ~ \\ \hline
        \textbf{FREEDOM\cite{zhou2023tale}} & 2023 & MM & \checkmark & ~ & ~ & ~ & ~ & ~ & ~ & ~ & ~ & ~ & ~ & ~ & ~ & ~ & \checkmark & ~ \\ \hline
        \textbf{MGCN\cite{yu2023multi}} & 2023 & MM & \checkmark & ~ & ~ & ~ & ~ & ~ & ~ & ~ & ~ & ~ & ~ & ~ & ~ & ~ & \checkmark & ~ \\ \hline
        \textbf{PaInvRL\cite{huang2023pareto}} & 2023 & MM & ~ & ~ & ~ & ~ & \checkmark & ~ & ~ & ~ & ~ & ~ & ~ & ~ & ~ & ~ & ~ & \checkmark \\ \hline
        \textbf{DRAGON\cite{zhou2023enhancing}} & 2023 & ECAI & \checkmark & ~ & ~ & ~ & ~ & ~ & ~ & ~ & ~ & ~ & ~ & ~ & ~ & ~ & \checkmark & ~ \\ \hline
        \textbf{LGMRec\cite{guo2024lgmrec}} & 2024 & AAAI & \checkmark & ~ & ~ & ~ & ~ & ~ & ~ & ~ & ~ & ~ & ~ & ~ & ~ & ~ & \checkmark & ~ \\ \hline
        \textbf{LLMRec\cite{wei2024llmrec}} & 2024 & WSDM & \checkmark & ~ & ~ & ~ & ~ & ~ & ~ & ~ & ~ & ~ & ~ & ~ & \checkmark & ~ & ~ & ~ \\ \hline
        \textbf{PromptMM\cite{wei2024promptmm}} & 2024 & WWW & \checkmark & ~ & ~ & ~ & ~ & ~ & ~ & ~ & ~ & ~ & ~ & ~ & \checkmark & ~ & ~ & ~ \\ \hline
        \textbf{MCDRec\cite{ma2024multimodal}} & 2024 & WWW & \checkmark & ~ & ~ & ~ & ~ & ~ & ~ & ~ & ~ & ~ & ~ & ~ & ~ & ~ & \checkmark & ~ \\ \hline
        \textbf{DiffMM\cite{jiang2024diffmm}} & 2024 & MM & \checkmark & ~ & ~ & ~ & ~ & ~ & ~ & ~ & ~ & ~ & ~ & ~ & \checkmark & ~ & ~ & ~ \\ \hline
        \textbf{SOIL\cite{su2024soil}} & 2024 & MM & \checkmark & ~ & ~ & ~ & ~ & ~ & ~ & ~ & ~ & ~ & ~ & ~ & ~ & ~ & \checkmark & ~ \\ \hline
        \textbf{CKD\cite{zhang2024modality}} & 2024 & MM & \checkmark & ~ & ~ & ~ & ~ & ~ & ~ & ~ & ~ & ~ & ~ & ~ & ~ & ~ & \checkmark & ~ \\ \hline
        \textbf{GUME\cite{lin2024gume}} & 2024 & CIKM & \checkmark & ~ & ~ & ~ & ~ & ~ & ~ & ~ & ~ & ~ & ~ & ~ & ~ & ~ & \checkmark & ~ \\ \hline
        \textbf{POWERec\cite{dong2024prompt}} & 2024 & INFFUS & \checkmark & ~ & ~ & ~ & ~ & ~ & ~ & ~ & ~ & ~ & ~ & ~ & ~ & ~ & \checkmark & ~ \\ \hline
        \textbf{DGVAE\cite{zhou2024disentangled}} & 2024 & TMM & \checkmark & ~ & ~ & ~ & ~ & ~ & ~ & ~ & ~ & ~ & ~ & ~ & ~ & ~ & \checkmark & ~ \\ \hline
        \textbf{VMoSE\cite{yi2024variational}} & 2024 & TMM & \checkmark & ~ & ~ & ~ & ~ & ~ & ~ & ~ & ~ & ~ & ~ & ~ & \checkmark & ~ & ~ & ~ \\ \hline
        \textbf{SAND\cite{he2024boosting}} & 2024 & arXiv & \checkmark & ~ & ~ & ~ & ~ & ~ & ~ & ~ & ~ & ~ & ~ & ~ & ~ & ~ & \checkmark & ~ \\ \hline
        \textbf{MENTOR\cite{xu2024mentor}} & 2025 & AAAI & \checkmark & ~ & ~ & ~ & ~ & ~ & ~ & ~ & ~ & ~ & ~ & ~ & ~ & ~ & \checkmark & ~ \\ 
        \textbf{SMORE\cite{ong2024spectrum}} & 2025 & WSDM & \checkmark & ~ & ~ & ~ & ~ & ~ & ~ & ~ & ~ & ~ & ~ & ~ & ~ & ~ & \checkmark & ~ \\ 
        \whline
    \end{tabular}
    \label{tab: Feature Extraction}
\end{table*}
We have summarized feature extraction in visual and textual modalities in Table~\ref{tab: Feature Extraction} for advanced MRS methods in recent years. Within the visual domain, early investigations predominantly utilized convolutional architectures such as CNNs, along with specific models like VGG \cite{simonyan2014very}, Inception \cite{szegedy2015going}, Caffe \cite{jia2014caffe}, and ResNet \cite{he2016deep}, which have demonstrated remarkable efficacy in handling various visual recognition tasks. These models are prized for their deep learning capabilities, which allow for the extraction of high-level, complex features from raw images. Moreover, in the textual domain, a diverse array of techniques has been employed for feature extraction. These include traditional methods like TF-IDF \cite{salton1988term}, as well as more sophisticated neural network approaches such as GRU \cite{cho2014properties}, PV-DM (PV-DBOW) \cite{koren2009bellkor}, and Glove \cite{pennington2014glove}. The introduction of attention mechanisms, along with models like BERT \cite{devlin2018bert}, Word2Vec \cite{mikolov2013distributed}, Sentence-Transformer \cite{reimers2019sentence}, and Sentence2Vec \cite{arora2017simple}, has further revolutionized the ability to understand and process language by allowing for contextually enriched text representations.

As the field progresses, the MMRec\footnote[1]{MMRec: https://github.com/enoche/MMRec.git} open-source framework exemplifies this trend by standardizing feature extraction methods for both visual and textual modalities, thereby facilitating a more controlled and reproducible experimental settings. Furthermore, data quality poses challenges, such as some corrupted and missing images in the Amazon datasets. Recent approaches have thus shifted towards utilizing pre-provided features in the visual modality to circumvent the issues related to manual processing. Similarly, in the textual modality, reliance on pre-trained models like BERT \cite{wei2023multi,yang2023modal,wei2024llmrec,wei2024promptmm,jiang2024diffmm,yi2024variational} or Sentence-Transformer \cite{zhang2021mining,mu2022learning,liu2022disentangled,zhou2023bootstrap,zhou2023tale,yu2023multi,zhou2023enhancing,guo2024lgmrec,ma2024multimodal,su2024soil,jiang2024diffmm,lin2024gume,dong2024prompt,zhou2024disentangled,he2024boosting,xu2024mentor} has become commonplace. These models provide a robust foundation for feature extraction, leveraging vast pre-existing knowledge bases to enhance the accuracy and depth of textual analysis.

Besides, not all studies have engaged both visual and textual modalities comprehensively. Several works have predominantly focused on the visual aspect as a form of auxiliary information. For instance, VBPR \cite{he2016vbpr} and DVBPR \cite{kang2017visually} underscore the significance of visual features in enhancing recommendation systems, while neglecting textual data. Similarly, VMCF \cite{park2017also} and ACF \cite{chen2017attentive} incorporate visual information to refine the accuracy of their models, yet they do not integrate textual insights which could potentially enrich the contextual understanding of the data. On the adversarial front, AMR \cite{tang2019adversarial} leverages visual modality to bolster the robustness of their models against adversarial attacks, yet the textual modality remains unexplored. Conversely, ADDVAE \cite{tran2022aligning} focuses exclusively on textual modality, thereby providing a nuanced understanding of textual data but omitting the rich, descriptive power of visual information. 

\section{Encoder}
\label{sec: Encoder}
\begin{figure*}
    \centering
    \includegraphics[width=1\linewidth]{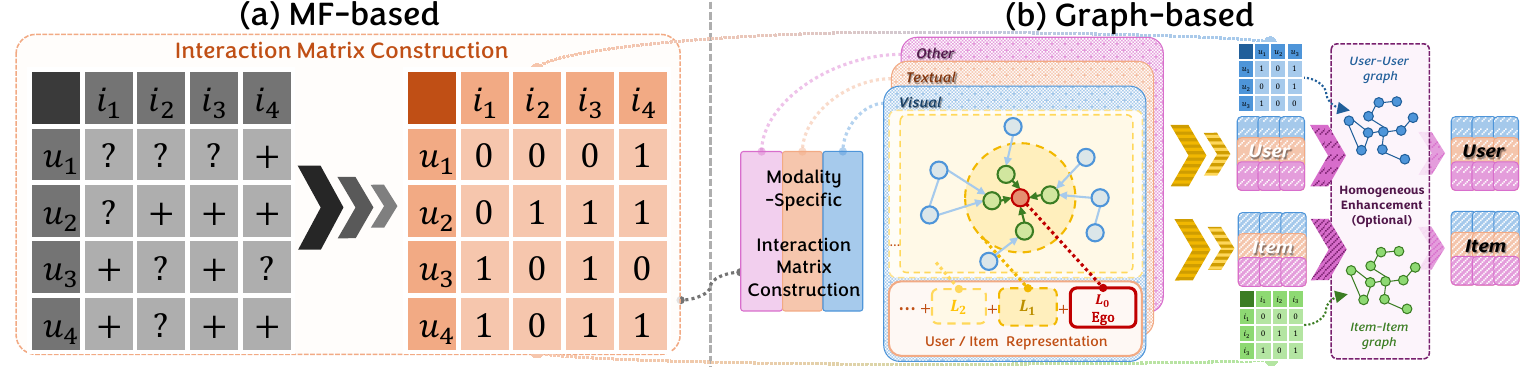}
    \caption{The illustration of two types of encoders.}
    \label{fig:Encoder}
\end{figure*}

In recommender systems, an encoder is typically used to extract a feature representation of a user or item. Encoders can be many types of models, from simple linear models to complex deep neural networks. The main purpose is to convert raw data (e.g., user behavioral data, item attributes, etc.) into a fixed-size embedding that captures the core features of the input data. However, in multimodal recommendation, the encoder plays the same role, but due to the greater variety of features, more diverse and specially designed encoders have been proposed to better utilize the multimodal information and user interaction data for more accurate recommendations. From a technical point of view, we roughly categorize all the encoders into MF-based encoders and Graph-based encoders. We depict these two types of encoders in Figure~\ref{fig:Encoder}.

\textbf{Preliminary 1:}  Let a set of users $\mathcal{U} \in \mathbb{R}^{\mathcal{|U|} \times d}$ and a set of items $\mathcal{I} \in \mathbb{R}^{\mathcal{|I|} \times d}$, where $d$ is the hidden dimension. We use $\mathbf{R} = [r_{u,i}]^{|\mathcal{U}| \times |\mathcal{I}|}$ to denote the user-item interaction matrix. For Graph-based encoder $\mathcal{G}$ = $(\mathcal{V}, \mathcal{E})$ be a given graph with node set $\mathcal{V}$ and edge set $\mathcal{E}$, where $|\mathcal{V}|$ = $|\mathcal{U}|$ + $|\mathcal{I}|$. $\mathbf{U}$ and $\mathbf{I}$ denote the hidden embeddings for users and items, respectively. $\mathbf{E} = \{\mathbf{U}|\mathbf{I}\}$ denotes the embedding for $\mathcal{V}$.

For the MRS scenario, we need to restate a more comprehensive preliminary.

\textbf{Preliminary 2:} Restate item sets $\mathcal{I}_m \in \mathbb{R}^{\mathcal{|I|} \times d_m}$, where $m \in \mathcal{M}$, $\mathcal{M}$ is the set of modalities, and $d_m$ is hidden dimension for modality $m$. Besides, $\mathcal{G}_m$ = $(\mathcal{V}_m, \mathcal{E})$ be a given graph with node set $\mathcal{V}_m$ and edge set $\mathcal{E}$, where $|\mathcal{V}_m|$ = $|\mathcal{U}|$ + $|\mathcal{I}_m|$. $\mathbf{I}_m$ denotes the items hidden embeddings for modality $m$. $\mathbf{E}_m = \{\mathbf{U}|\mathbf{I}_m\}$ denotes the embedding for $\mathcal{V}_m$.

\subsection{MF-based Encoder} The core idea of MF-based encoders is to decompose the user-item rating matrix $\mathbf{R}$ into two low-rank hidden embeddings $\mathbf{U}$ and $\mathbf{I}$. The approximation of the rating matrix $\mathbf{R}$ can be expressed as:
\begin{equation}
\label{eq: mf}
    \mathbf{R} \approx \hat{\mathbf{R}} = \mathbf{U} \mathbf{I}^T,
\end{equation}
where $^T$ means the transpose operation for the matrix. The loss function can be defined as:
\begin{equation}
\min _{\mathbf{U}, \mathbf{I}}\|\mathbf{R}-\hat{\mathbf{R}}\|_F^2+\lambda(\|\mathbf{E}\|_F^2),
\end{equation}
where $\|\cdot\|_F$ means Frobenius norm and $\lambda$ is the regularization parameter to control overfitting.

Due to the diversified features of the item, we categorize the strategies into a unified MF-based encoder and multiple MF-based encoders. The design of both relies on the choice of multimodal fusion, which we will discuss in detail in Section~\ref{sec: Multimodal Fusion}. Roughly speaking, the approximation of the rating matrix $\mathbf{R}$ for MRS can be expressed as:
\begin{equation}
    \mathbf{R} \approx \hat{\mathbf{R}} = \mathbf{U} \mathbf{I}^T, \quad \mathbf{I} = \operatorname{Aggr}(\mathbf{I}_m),
\end{equation}
\begin{equation}
    \mathbf{R} \approx \hat{\mathbf{R}} = \operatorname{Aggr}(\mathbf{U} \mathbf{I}_m^T),
\end{equation}
where $\operatorname{Aggr}(\cdot)$ denotes multimodal fusion. The loss function can be defined as:
\begin{equation}
\label{eq: gcn}
\min _{\mathbf{U}, \mathbf{I}}\|\mathbf{R}-\hat{\mathbf{R}}\|_F^2+\lambda(\sum_{m \in M}\|\mathbf{E}_m\|_F^2).
\end{equation}

\subsection{Graph-based Encoder} The core idea of Graph-based encoders is to use the features of nodes and the structural information of the graph to learn the representation of nodes. We introduce the commonly used graph-based encoder paradigm Graph Convolution Network (GCN) \cite{kipf2017semi}. For a graph $\mathcal{G}$ and its adjacency matrix $\mathbf{A}$, a layer propagation of GCN can be expressed as:
\begin{equation}
\mathbf{E}^{(l)}=\sigma(\mathbf{\tilde{D}}^{-\frac{1}{2}} \tilde{\mathbf{A}} \tilde{\mathbf{D}}^{-\frac{1}{2}} \mathbf{E}^{(l-1)} \mathbf{W}^{(l-1)}),
\end{equation}
where $\mathbf{E}^{(l)}$ is the $l$ layer hidden embeddings for $\mathcal{V}$, $\tilde{\mathbf{A}}$ $=$ $\mathbf{A} + \mathbf{I}$. $\tilde{\mathbf{D}}$ is the degree matrix of $\tilde{\mathbf{A}}$, where $\tilde{\mathbf{D}}_{i i}=\sum_j \tilde{\mathbf{A}}_{ij}$. $\mathbf{W}^{(l-1)}$ is the weight matrix for $l$$-$$1$ layer and $\sigma(\cdot)$ is the active function. In the recommendation domain, LightGCN \cite{he2020lightgcn} proves that weight matrices and active functions in GCNs are useless and even increase training difficulty. To this end, a widely-used simplified GCN can be expressed as:
\begin{equation}
\mathbf{E}^{(l)}=\mathbf{\tilde{D}}^{-\frac{1}{2}} \tilde{\mathbf{A}} \tilde{\mathbf{D}}^{-\frac{1}{2}} \mathbf{E}^{(l-1)},
\end{equation}
Then it stacks multiple layers by $\bar{\mathbf{E}} = \operatorname{Stack}_{l \in L}(\mathbf{E}^{(l)})$, where $L$ is the total layer number of GCN. We simply define this entire Graph-based Encoder as $\bar{\mathbf{E}} = \operatorname{GCN}(\mathbf{U},\mathbf{I})$. The entire representation $\bar{\mathbf{E}}$ can be split into user and item parts by $\bar{\mathbf{U}}, \bar{\mathbf{I}} = \operatorname{Sp}(\bar{\mathbf{E}})$. To better mine user and item representations, two homogeneous type graphs, user-user and item-item, are proposed. A portion of Graph-based encoders will use either one or all of them to better learn the representations. Homogeneous graphs first retain the top-$k$ items/users by similarity:
\begin{equation}
\mathbf{S}^I_{i, i^{\prime}}= \begin{cases}1, & \mathbf{S}^I_{i, i^{\prime}} \in \text {top-} k(\mathbf{S}^I_{i, i^{\prime}}) \\ 0, & \text {otherwise}\end{cases},
\end{equation}
\begin{equation}
\mathbf{S}^U_{u, u^{\prime}}= \begin{cases}1, & \mathbf{S}^U_{u, u^{\prime}} \in \text {top-} k(\mathbf{S}^U_{u, u^{\prime}}) \\ 0, & \text {otherwise}\end{cases}.
\end{equation}
To mitigate the issue of significant disparities in the feature distributions between vertices with high degrees and those with low degrees within graph-structured data, it is a common practice to apply symmetric normalization to the adjacency matrix, denoted as $\mathbf{S}^U=(\mathbf{D}^U)^{-1 / 2} \mathbf{S}^U(\mathbf{D}^U)^{-1 / 2}$ and $\mathbf{S}^I=(\mathbf{D}^I)^{-1 / 2} \mathbf{S}^I(\mathbf{D}^I)^{-1 / 2}$, 
where $\mathbf{D}^U$ and $\mathbf{D}^I$ represent the diagonal degree matrix of $\mathbf{S}^U$ and $\mathbf{S}^I$, respectively. This normalization process is crucial as it adjusts the influence of each vertex based on its connectivity, thereby preventing vertices with a higher degree from disproportionately dominating the feature representations. Then propagate $\mathbf{\bar{U}}$/$\mathbf{\bar{I}}$ through:
\begin{equation}
    \mathbf{\hat{U}} = (\mathbf{S}^U)^{L_u}\mathbf{\bar{U}}, \quad \mathbf{\hat{I}} = (\mathbf{S}^I)^{L_i}\mathbf{\bar{I}},
\end{equation}
where $L_u$ and $L_i$ are the layer number of the user-user graph and item-item graph, respectively. These two representations can optionally enhance user and item representations by $\mathbf{\tilde{U}} = \mathbf{\hat{U}} + \mathbf{\bar{U}}$ and  $\mathbf{\tilde{I}} = \mathbf{\hat{I}} + \mathbf{\bar{I}}$.

For the sake of simplicity, we define the composite Graph-based encoder including optional user-user and item-item graphs as:
\begin{equation}
    \mathbf{\tilde{U}},\mathbf{\tilde{I}} = \operatorname{C-GCN}(\mathbf{U},\mathbf{I}))
\end{equation}
The approximation of the rating matrix $\mathbf{R}$ for MRS can be expressed as:
\begin{equation}
\mathbf{R} \approx \mathbf{\hat{R}} = \mathbf{\tilde{U}}\mathbf{\tilde{I}}^T, \quad \mathbf{\tilde{U}},\mathbf{\tilde{I}} = \operatorname{C-GCN}(\mathbf{U},\mathbf{I})).
\end{equation}
The loss function can be defined as Eq~\ref{eq: mf}. Same as MF-based encoders, Graph-based encoders can also be categorized into a unified Graph-based encoder and multiple Graph-based encoders with different fusion strategies, detailed in Section~\ref{sec: Multimodal Fusion}. Roughly speaking, the approximation of the rating matrix $\mathbf{R}$ for MRS can be expressed as:
\begin{equation}
    \mathbf{R} \approx \hat{\mathbf{R}} = \mathbf{\tilde{U}}\mathbf{\tilde{I}}^T, \quad  \tilde{\mathbf{U}},\tilde{\mathbf{I}} = \operatorname{C-GCN}(\mathbf{U},\operatorname{Aggr}(\mathbf{I}_m)),
\end{equation}
\begin{equation}
    \mathbf{R} \approx \hat{\mathbf{R}} = \operatorname{Aggr}(\mathbf{\tilde{U}} \mathbf{\tilde{I}}_m^T), \quad \mathbf{\tilde{U}},\mathbf{\tilde{I}}_m = \operatorname{C-GCN}(\mathbf{U},\mathbf{I}_m).
\end{equation}
The loss function can be defined as Eq~\ref{eq: gcn}.

\subsection{Taxonomy}
\begin{figure*}
    \centering
    \includegraphics[width=1\linewidth]{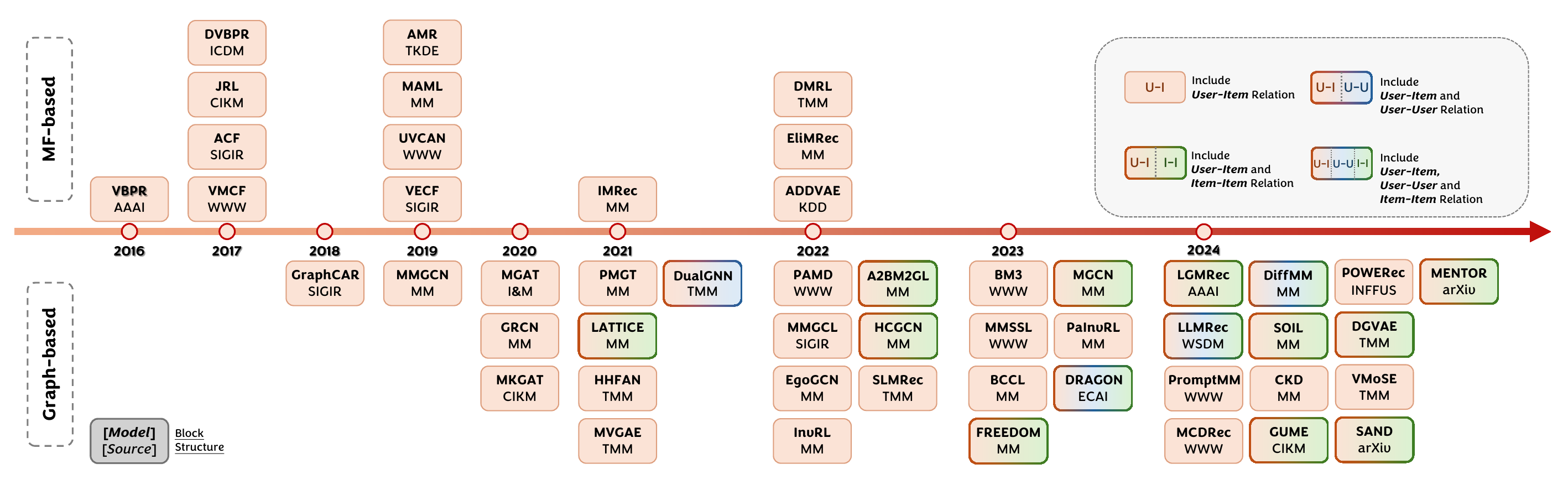}
    \caption{Taxonomy of Encoders.}
    \label{fig:Summary of Encoders}
\end{figure*}

Following this analysis, we categorize recent works in MRS into MF-based and Graph-based encoders, as summarized in Figure~\ref{fig:Summary of Encoders}. For Graph-based encoders, we delineate the types of relationships, including user-item (U-I), user-user (U-U), and item-item (I-I) relations. 

\textbf{MF-based Encoder:} VBPR \cite{he2016vbpr} integrates visual features extracted by a pre-trained deep CNN into MF for enhanced preference prediction. VMCF \cite{park2017also} builds a product-affinity network that incorporates visual appearance and inter-item relationships. ACF \cite{chen2017attentive} introduces item- and component-level attention mechanisms to better handle implicit feedback, marking a first in applying attention in CF. JRL \cite{zhang2017joint} and its extension, eJRL, employ multi-view machine learning to merge diverse information sources, improving top-N recommendations without needing to retrain for new data. DVBPR \cite{kang2017visually} and UVCAN \cite{liu2019user} enhance recommendations by integrating visual signals and employing co-attention mechanisms, respectively. MAML \cite{liu2019user2} and ADDVAE \cite{tran2022aligning} focus on modeling user preferences using concatenated textual and visual inputs through neural networks, with ADDVAE also exploring disentangled representations. AMR \cite{tang2019adversarial} addresses vulnerabilities in MRS using adversarial learning to create more robust models. IMRec aligns recommendation processes with user reading habits by focusing on local news details. EliMRec \cite{liu2022elimrec} and DMRL \cite{liu2022disentangled} utilize causal inference and disentangled representations, respectively, to reduce biases and better capture independent modal factors.

\textbf{Graph-based Encoder:} GraphCAR \cite{xu2018graphcar} combines multimedia content with the traditional CF method. MMGCN \cite{wei2019mmgcn} employs GCNs to learn representations for each modality, which are then fused with ID embeddings to form the final item representations. Building on the MMGCN framework, MGAT \cite{tao2020mgat} uses standard GCN aggregation and a similar fusion method for combining results. GRCN \cite{wei2020graph} refines user-item interaction graphs by identifying and cutting noise edges. MKGAT \cite{sun2020multi} constructs a multimodal knowledge graph with an entity-based approach, using specialized encoders for different data types and an attention layer for effective information aggregation from neighboring entities. PMGT \cite{liu2021pre} is a pre-trained model that utilizes fused multimodal features and interactions, employing an attention mechanism to derive multimodal embeddings. These embeddings are then enhanced with position and role-based embeddings to initialize node embeddings for pre-training and subsequent downstream tasks. DualGNN \cite{wang2021dualgnn} introduces a user-user graph to uncover hidden preference patterns among users. LATTICE \cite{zhang2021mining} develops an item-item graph to detect semantically correlated signals among items. HHFAN \cite{cai2021heterogeneous} develops a heterogeneous graph incorporating user, item, and multimodal information, and uses random walks to sample neighbors based on node type. It employs a Fully Connected (FC) layer to unify various node vectors into a single space and uses LSTM for aggregating embeddings of the same node type within the intra-type feature aggregation network. MVGAE \cite{yi2021multi} is a multimodal variational graph auto-encoder model that utilizes the modality-specific variational encoder to learn the node representation. PAMD \cite{han2022modality} employs a disentangled encoder to separate common and unique characteristics of objects into respective representations, and uses contrastive learning for cross-modality alignment of these representations. MMGCL \cite{yi2022multi} integrates self-supervised learning with graph-based approaches for micro-video recommendation, featuring an innovative negative sampling method that highlights inter-modality relationships. EgoGCN \cite{chen2022breaking} introduces an effective graph fusion method, which is not confined to unimodal graph information propagation but aggregates informative inter-modal messages from neighboring nodes. InvRL \cite{du2022invariant} addresses spurious correlations in multimedia recommendations by learning stable item representations across diverse environments. A2BM2GL \cite{cai2022adaptive} integrates collaborative and semantic representation learning to effectively model nodes and video features. An anti-bottleneck module with attention mechanisms enhances node relationship expressiveness. Additionally, an adaptive recommendation loss dynamically adjusts to user preference variations, improving item recommendation accuracy. HCGCN \cite{mu2022learning} extends the LATTICE framework by utilizing co-clustering and item-clustering losses to refine user-item preference feedback and adjust modality importance. SLMRec \cite{tao2022self} proposes a self-supervised learning framework for multimodal recommendations, establishing a node self-discrimination task to reveal hidden multimodal patterns of items. BM3 \cite{zhou2023bootstrap} simplifies SLMRec by replacing the random negative example sampling mechanism with a dropout strategy. MMSSL \cite{wei2023multi} designs a modality-aware interactive structure learning paradigm via adversarial perturbations, and proposes a cross-modal comparative learning method to disentangle the common and specific features among modalities.BCCL \cite{yang2023modal} integrates a bias constraint module for data augmentation, a modal awareness module, and a sparse enhancement module to collaboratively produce high-quality samples. FREEDOM \cite{zhou2023tale} refines LATTICE by freezing the item-item graph and reducing noise in the user-item graph. MGCN \cite{yu2023multi} purifies modal features using item behavior information to reduce noise contamination and models modal preferences based on user behavior. PaInvRL \cite{huang2023pareto} adaptively balances ERM (Empirical Risk Minimization) and IRM (Invariant Risk Minimization) losses to achieve Pareto-optimal solutions, effectively enhancing model performance by minimizing losses for Pareto optimality. DRAGON \cite{zhou2023enhancing} learns the dual representations of users and items by constructing homogeneous and heterogeneous graphs. LGMRec \cite{guo2024lgmrec} integrates local embeddings, which capture local topological nuances, with global embeddings, which consider hypergraph dependencies. LLMRec \cite{wei2024llmrec} employs three simple yet effective LLM-based graph augmentation strategies to enhance recommendation performance. PromptMM \cite{wei2024promptmm} enhances knowledge distillation by disentangling collaborative relationships to enable augmented distillation. MCDRec \cite{ma2024multimodal} integrates modal perceptual features with collaborative information to improve item representation and employs diffusion-aware representation to denoise the user-item interaction graph. DiffMM \cite{jiang2024diffmm} introduces a well-designed modality-aware graph diffusion model to improve modality-aware user representation learning. SOIL \cite{su2024soil} exploits candidate items from the perspective of constructing interest-aware graphs. CKD \cite{zhang2024modality} aims to solve the modal imbalance problem and make the best use of all modalities. MENTOR \cite{xu2024mentor} leverage aligned modalities while preserving interaction information with multi-level cross-modal alignment. GUME \cite{lin2024gume} achieves outstanding performance in long-tail scenarios by combining MGCN and MENTOR. POWERec \cite{dong2024prompt} leverages prompt learning to model modality-specific interests. DGVAE \cite{zhou2024disentangled} utilizes GCNs to encode and disentangle ratings and multimodal information, learning item representations from an item-item graph and enhancing interpretability by projecting multimodal data into text. VMoSE \cite{yi2024variational} enhances robustness by adaptive sampling and fusing noisy multi-modal signals based on uncertainty estimates. SAND \cite{he2024boosting} aligns modal-generic representations efficiently without negative sampling and distinctly separates modal-unique representations to preserve modality-independent information.

% \input{Tab/Encoder}

% \red{Todo}
\section{Multimodal Fusion}
\label{sec: Multimodal Fusion}

\begin{table*}[t]
    \centering
    \caption{Simplified summary of fusion strategy. We simplified express them with only two modalities visual and textual. $E_v$ and $E_t$ denote representations for visual and textual modalities, respectively.}
    % \resizebox{\linewidth}{!}{
    \begin{tabular}{|m{2cm}<{\centering}|m{5cm}<{\centering}|m{5cm}<{\centering}|}
    \hline
         \textbf{Fusion Strategy} &\textbf{Concatenation}&  \textbf{Element-wise}\\
         \hline
         \textbf{Heuristic} & $E_v | E_t$ & $E_v + E_t$ \\ \hline
         \textbf{Attentive} & $\alpha_v E_v | \alpha_t E_t$ & $\alpha_v E_v + \alpha_t E_t$ \\ \hline
    \end{tabular}
    % }
    \\
    $|$ denotes concatenation operation. $\alpha_v$ and $\alpha_t$ denote learnable weights for each modality.
    \label{tab: Fusion Strategy}
\end{table*}

Multimodal fusion represents a critical investigation area in MRS. The effectiveness of multimodal integration is significantly influenced by the timing of when different modalities are fused, as well as the strategies employed to execute this fusion. Therefore, it becomes essential to systematically analyze and categorize existing works based on these two perspectives: timing and strategy. 

\begin{figure}[h]
    \centering
    \includegraphics[width=1\linewidth]{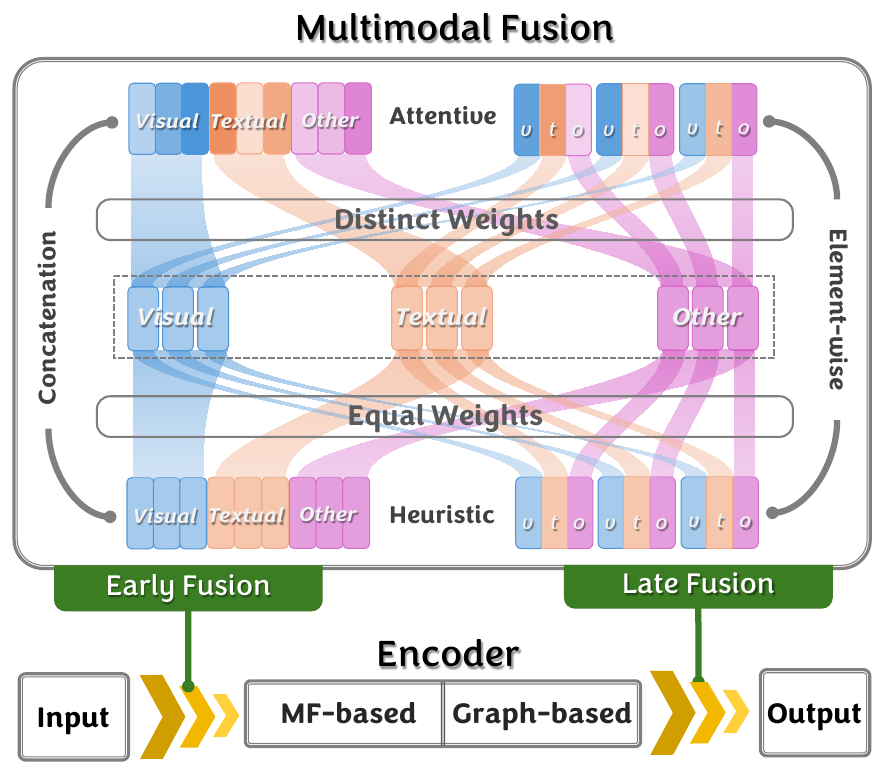}
    \caption{The illustration of Multimodal Fusion.}
    \label{fig:Multimodal Fusion}
\end{figure}

\subsection{Timing Perspective}
We first categorize all recent works in MRS from the timing perspective. Specifically, it can be decided into \textbf{Early fusion} and \textbf{Late fusion} strategies, respectively. Formally,
\begin{equation}
\label{eq:early}
    \text{Early fusion: }\mathbf{\bar{E}} = \operatorname{Encoder}(\mathbf{U},\operatorname{Aggr}(\mathbf{I}_m)),
\end{equation}
\begin{equation}
\label{eq:late}
   \text{Late fusion: }\mathbf{\bar{E}} = \operatorname{Aggr}(\operatorname{Encoder}(\mathbf{U},\mathbf{I}_m)),
\end{equation}
where we utilize the same preliminaries as detailed in Section~\ref{sec: Encoder}. $\operatorname{Encoder}$ denotes various encoders. Early fusion denotes fusing all modalities before graph message propagation and aggregation. Late fusion denotes fusing all modalities after graph message propagation and aggregation. Although both early and late fusion timings demonstrate commendable performance in various scenarios, each approach exhibits distinct limitations. Specifically, early fusion tends to integrate modalities at the early stage, which can result in modal-specific features not being fully exploited due to premature integration. On the other hand, the late fusion strategy, which combines modalities at a later stage, faces challenges in fully capturing and leveraging the correlations among different modalities. It might limit the model's ability to extract the correlation among different modalities, potentially reducing the overall effectiveness of leveraging multimodal information. 

\subsection{Strategy Perspective}
From the strategy perspective, all recent works can be fine-grained categories from two dimensions, \textbf{Element-wise or Concatenation} and \textbf{Attentive or Heuristic}. Many works in MRS have introduced subtle adjustments to the fusion strategy. Despite these variations, it is commonly feasible to categorize these approaches in a broad manner based on the underlying motivations driving their use, as delineated in Table~\ref{tab: Fusion Strategy}.

The method of element-wise fusion provides a more profound integration of different modalities compared to the concatenation approach. However, this deeper integration might inadvertently amplify the inherent noise present within the modality data. On the other hand, the attentive approach, as opposed to heuristic methods, offers a more dynamic allocation of modality weights, allowing for a more adaptive and responsive handling of input features based on their relevance to the specific task. Despite its advantages, the attentive mechanism incurs significantly higher computational costs and increases the complexity of the training phase. 

The fusion strategy often interacts synergistically with the timing of fusion, thereby influencing the model performance significantly. Recognizing this interplay, we conduct a comprehensive summary of the recent strategies and timing of modality fusion within MRS, as shown in Figure~\ref{fig:Multimodal Fusion}. Our objective is to provide researchers with a clearer perspective on how these perspectives coalesce to influence model performance.

\begin{figure}[h]
    \centering
    \includegraphics[width=0.9\linewidth]{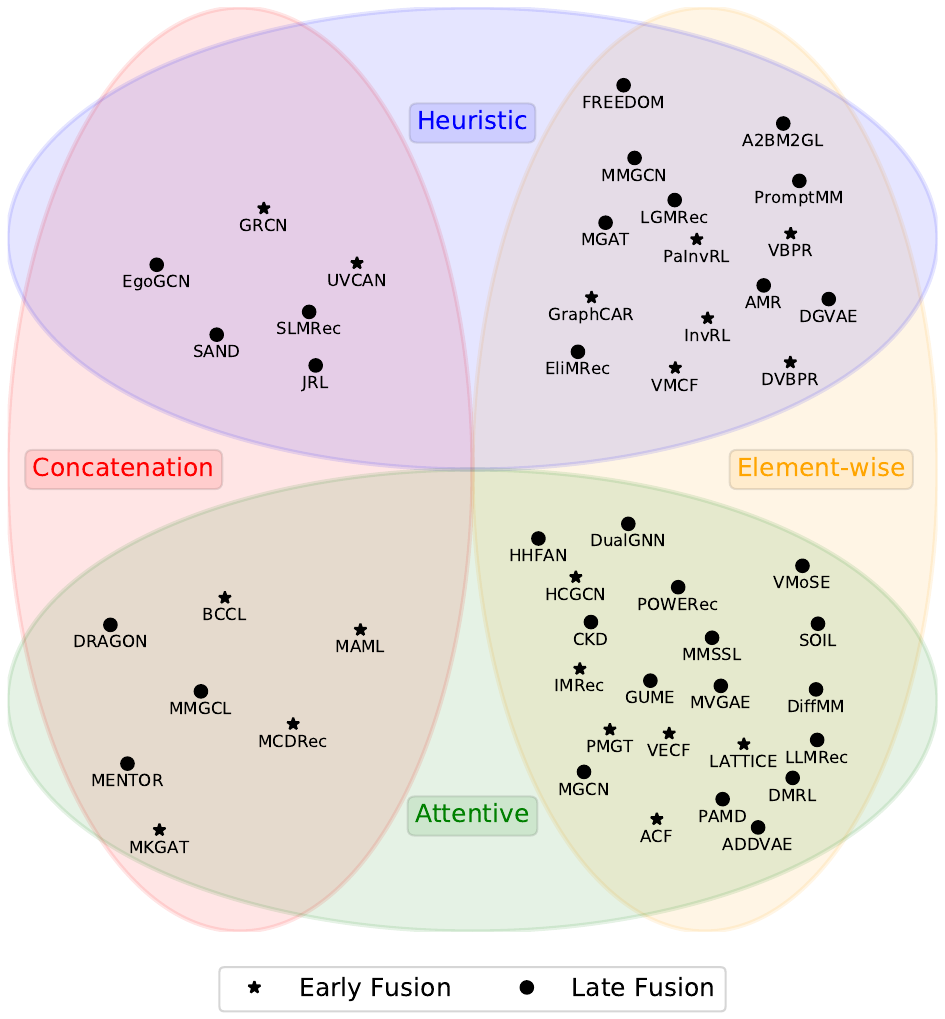}
    % \vskip -0.1in
    \caption{Taxonomy of Multimodal Fusion. five-pointed star and circle symbols denote early fusion and late fusion, respectively.}
    \label{fig:category}
    % \vskip -0.2in
\end{figure}

\subsection{Taxonomy}
Following the above analysis, we categorize recent works from both fusion timing and fusion strategy perspectives, as summarized in Figure~\ref{fig:category}. We further provide a specific analysis of all recent MRS works from a fusion timing perspective. 

\textbf{Early Fusion:} VBPR \cite{he2016vbpr} and GraphCAR \cite{xu2018graphcar} directly utilize visual features within MF frameworks. In an extension to this approach, VMCF \cite{park2017also} introduces visual matrix co-factorization. DVBPR \cite{kang2017visually} leverages a deeper neural network to enhance the representation of visual data. Attention mechanisms have been increasingly applied to dynamically allocate weights across different modalities. Works such as ACF \cite{chen2017attentive}, VECF \cite{chen2019personalized}, UVCAN \cite{liu2019user}, MAML \cite{liu2019user2}, GRCN \cite{wei2020graph}, IMRec \cite{xun2021we}, PMGT \cite{liu2021pre}, BCCL \cite{yang2023modal}, and MCDRec \cite{ma2024multimodal} utilize attention mechanisms to effectively allocate weights for different modalities. MKGAT \cite{sun2020multi} constructs a multimodal knowledge graph that fuses modal features to enhance recommendation systems. More recently, LATTICE \cite{zhang2021mining} exploits raw features to construct item-item graphs for each modality and dynamically fuses weights using an attention mechanism to form modality-specific graphs. Moreover, HCGCN \cite{mu2022learning} employs a multimodal item-item graph to enhance the user-item graph, enabling the discovery of user preferences within similar items and facilitating simultaneous clustering. 

\textbf{Late Fusion:} Models such as JRL \cite{zhang2017joint}, MMGCN \cite{wei2019mmgcn}, DualGNN \cite{wang2021dualgnn}, MMGCL \cite{yi2022multi}, EgoGCN \cite{chen2022breaking}, FREEDOM \cite{zhou2023tale}, MGCN \cite{yu2023multi}, DRAGON \cite{zhou2023enhancing}, LGMRec \cite{guo2024lgmrec}, DiffMM \cite{jiang2024diffmm}, SOIL \cite{su2024soil}, and POWERec \cite{dong2024prompt} first independently processing different modalities and then combining them effectively to enhance prediction accuracy. Moreover, HHFAN \cite{cai2021heterogeneous} employs self-attention-aware neural networks to integrate information from all modalities, enhancing the model's ability to focus on more informative features. Attention mechanisms are applied in PAMD \cite{han2022modality}, A2BM2GL \cite{cai2022adaptive}, and LLMRec \cite{wei2024llmrec}, which dynamically adjust the influence of each modality on the final prediction, providing a flexible and context-sensitive fusion strategy. Similarly, MVGAE \cite{yi2021multi} uses a product-of-experts framework to harmonize modality-specific distributions, ensuring an effective fusion of modal inputs. Furthermore, some methods focus on leveraging auxiliary modality alignment tasks to better integrate modalities. ADDVAE \cite{tran2022aligning}, MMSSL \cite{wei2023multi}, PromptMM \cite{wei2024promptmm}, GUME \cite{lin2024gume}, DGVAE \cite{zhou2024disentangled}, SAND \cite{he2024boosting}, VMoSE \cite{yi2024variational}, and MENTOR \cite{xu2024mentor} all incorporate auxiliary tasks to help modality fusion. Additionally, EliMRec\cite{liu2022elimrec} integrate modalities from a causal perspective, and CKD \cite{zhang2024modality} applies the Average Treatment Effect (ATE) strategy for modality fusion, aiming to quantify the impact of each modality on the recommendation outcome effectively. SMORE \cite{ong2024spectrum} projects the multi-modal features into the frequency domain and leverages the spectral space for fusion.

Since the goal of each fusion strategy is relatively fixed, we will not go into too much detail from the perspective of the fusion strategy.
\section{Loss Function}

\begin{figure*}
    \centering
    \includegraphics[width=1\linewidth]{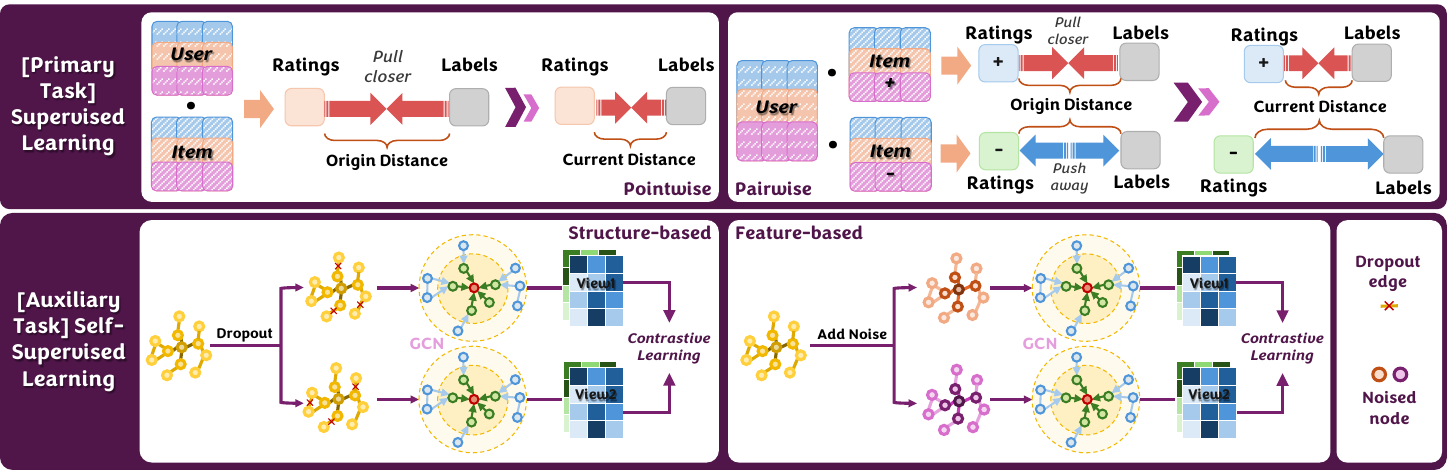}
    \caption{The illustration of Loss Functions.}
    \label{fig:Loss Functions}
\end{figure*}

\label{sec: Loss Function}

MRS utilizes loss functions consisting of two key types: primary and auxiliary tasks, as shown in Figure~\ref{fig:Loss Functions}. The primary tasks involve supervised learning, where clearly defined labels guide the model’s learning, ensuring accurate predictions from labeled data. Conversely, the auxiliary tasks employ self-supervised learning (SSL) \cite{liu2021self}, which generates supervisory signals from the data's inherent structure, independent of external labels. This method enables effective use of unlabeled data, allowing the system to extract meaningful representations and maintain accuracy, even when labeled data is sparse. We introduce these two types of loss functions in detail.

\subsection{Supervised Learning}

\textbf{Preliminary:} In the RS scenario, we simplified define the recommendation model as $f(\cdot)$, it can predict the score between user $u$ and item $i$ by:
\begin{equation}
    y_{u,i} = f(\mathbf{R}, u, i),
\end{equation}
where $\mathbf{R} = [r_{u,i}]^{|\mathcal{U}| \times |\mathcal{I}|}$ represents the user-item interaction matrix. In the MRS scenario, the multimodal recommendation model will be expanded as $f_m(\cdot)$, it can predict the score between user $u$ and item $i$ with multimodal information by:
\begin{equation}
    y_{u,i} = f_m(\mathbf{R}, u, i, \mathbf{M}),
\end{equation}
where $\mathbf{M}$ incorporates external item attributes with different modalities.

The supervised learning task is to force the predicted score between any user $u$ and item $i$ closer to the ground label. It can be divided into \textbf{Pointwise Loss} and \textbf{Pairwise Loss}. 

For Pointwise Loss, we introduce two common loss functions that are also widely used in many machine learning downstream tasks, such as Computer Vision and Natural Language Processing. 1) Mean Squared Error (MSE) \cite{marmolin1986subjective} and 2) Cross-Entropy Loss (CE) \cite{lecun2015deep}. 

MSE can be defined as:
\begin{equation}
    \frac{1}{|\mathcal{D}|} \sum_{(u,i) \in \mathcal{D}}(r_{u,i}-y_{u,i})^2,
\end{equation}
where $\mathcal{D}$ denotes the training set. This loss function pulls the predicted score for interacted user-item pairs more approximately to 1, and non-interacted user-item pairs more approximately to 0.

CE can be defined as:
\begin{equation}
-\frac{1}{|\mathcal{D}|} \sum_{(u,i) \in \mathcal{D}}[r_{u,i} \log (y_{u,i})+(1-r_{u,i}) \log (1-y_{u,i})].
\end{equation}
Utilizing CE in the RS (MRS) fields treats the prediction function as a two-fold classification task, classifying user-item pairs into interacted and non-interacted categories. This improves classification accuracy by minimizing the discrepancy between the predicted probability distribution and the true label distribution, allowing the model to better learn the probability distribution of classified labels. However, due to the natural sparsity of user-item interaction data, we need to value the scarce labeled data more, and thus further propose the Pairwise Loss.

For Pointwise Loss, the learning objective is to pull closer the positive pairs, while pushing away the negative pairs. We introduce two common loss functions that are also widely used loss function. 1) Bayesian Personalized Ranking (BPR) \cite{rendle2012bpr} and 2) Hinge Loss \cite{cortes1995support}.

BPR can be defined as:
\begin{equation}
    \sum_{(u, i^+, i^-) \in \mathcal{D}} - \log(\sigma(y_{u,i^+} - y_{u,i^-})),
\end{equation}
where $y_{u,i^+}$ and $y_{u,i^-}$ are the ratings of user $u$ to the positive item $i^+$ and negative item $i^-$. $\sigma$ is the active function.

Hinge Loss can be defined as:
\begin{equation}
\max (0,1-r_{u,i} \cdot y_{u,i}).
\end{equation}
Hinge aims to not only correctly classify data points, but also to maximize the spacing of classification decision boundaries.

\subsection{Self-supervised Learning}
\textbf{Preliminary:} In the RS scenario, self-supervised learning first creates two different views for contrastive learning, we simply define this view creator as: 
\begin{equation}
    w = \mathcal{C}(\mathbf{R}, \mathbf{E}),
\end{equation}
where $\mathcal{C}(\cdot)$ is a view creator. $\mathbf{R}$ and $\mathbf{E}$ are use the same definition as Section~\ref{sec: Encoder}. In the MRS scenario, the view creator can be further defined as:
\begin{equation}
    w = \mathcal{C}(\mathbf{R}, \mathbf{E}, \mathbf{M}).
\end{equation}

The view creator can be divided into two types: 1) \textbf{Feature-based SSL} and 2) \textbf{Structure-based SSL}.

Feature-based SSL is designed to generate multiple views of the same data instance by perturbing the features. This approach is strategically employed to enhance the robustness of the learned representations. By introducing variations in the input features, the model is compelled to focus on the invariant aspects across these modifications, thereby acquiring a deeper, more generalized understanding of the user-item interaction. It can be simplify expressed as:
\begin{equation}
    \omega = \mathcal{C}_{feature}(\mathbf{R}, \mathbf{E}, \mathbf{M}) = (\mathbf{R}, \operatorname{Perturb}(\mathbf{E}), \mathbf{M}),
\end{equation}
where $\operatorname{Perturb}(\cdot)$ can be an MLP, feature dropout, adding random noise, etc. It is worth noting that different modalities can be naturally considered as two feature-based views. Applying feature-based SSL for two different modalities representation can be seen as modality alignments \cite{xu2024mentor}.

Structure-based SSL is designed to generate multiple views of the same data instance by perturbing the graph structures. This approach is employed to intricately capture the complex dependencies and interactions inherent within graph structures, which are pivotal in enhancing the robustness and performance of graph-based learning models. Such manipulations enable the learning algorithms to discern and generalize from the essential features of the data, potentially leading to more effective and insightful representations in domains where graph-based data is prevalent. It can be simplify expressed as:
\begin{equation}
    \omega = \mathcal{C}_{structure}(\mathbf{R}, \mathbf{E}, \mathbf{M}) = (\operatorname{Perturb}(\mathbf{R}), \mathbf{E}, \mathbf{M}),
\end{equation}
where $\operatorname{Perturb}(\cdot)$ can be an MLP, node/edge dropout, adding random noise, etc.

The self-supervised learning task is to force the generated views closer to enhance the representation ability of the MRS model. We introduce two widely used self-supervised loss functions. 1) InfoNCE \cite{oord2018representation} and 2) Jensen-Shannon divergence (JS) \cite{hjelm2018learning}.

InfoNCE is a variant of Noise Contrastive Estimation \cite{gutmann2010noise}, which gained wide adoption as a self-supervised learning loss function in RS. It can be expressed as:
\begin{equation}
\mathbb{E}[-\log \frac{\exp (f(\omega_i^{\prime}, \omega_i^{\prime \prime}))}{\sum_{\forall i, j} \exp (f(\omega_i^{\prime}, \omega_j^{\prime \prime}))}],
\end{equation}
where $f(\cdot)$ represents a critic function that calculates a score indicating the similarity between two views. The term $\exp f(\omega_i^{\prime}, \omega_i^{\prime \prime})$ corresponds to the score of positive pairs, while the term $\sum \exp (f(\omega_i^{\prime}, \omega_j^{\prime \prime}))$ encompasses both the numerator and the scores of all negative pairs.

In addition to using InfoNCE estimation for mutual information, the lower bound can also be estimated using the Jensen-Shannon (JS) divergence. The derived learning objective is akin to
combining InfoNCE with a standard binary cross-entropy loss \cite{xia2021self}, applied to positive pairs and negative pairs. 
\begin{equation}
\mathbb{E}[-\log \sigma(f(\omega_i^{\prime}, \omega_i^{\prime \prime}))]-\mathbb{E}[\log (1-\sigma(f(\omega_i^{\prime}, \omega_j^{\prime \prime})))],
\end{equation}
where $\sigma$ represents the sigmoid function used to normalize the output of the critic function. The main idea behind this optimization is to assign the label 1 to positive pairs and 0 to negative pairs, thereby increasing the predicted value for positive pairs and enhancing the similarity between them.

\begin{figure*}
    \centering
    \includegraphics[width=1\linewidth]{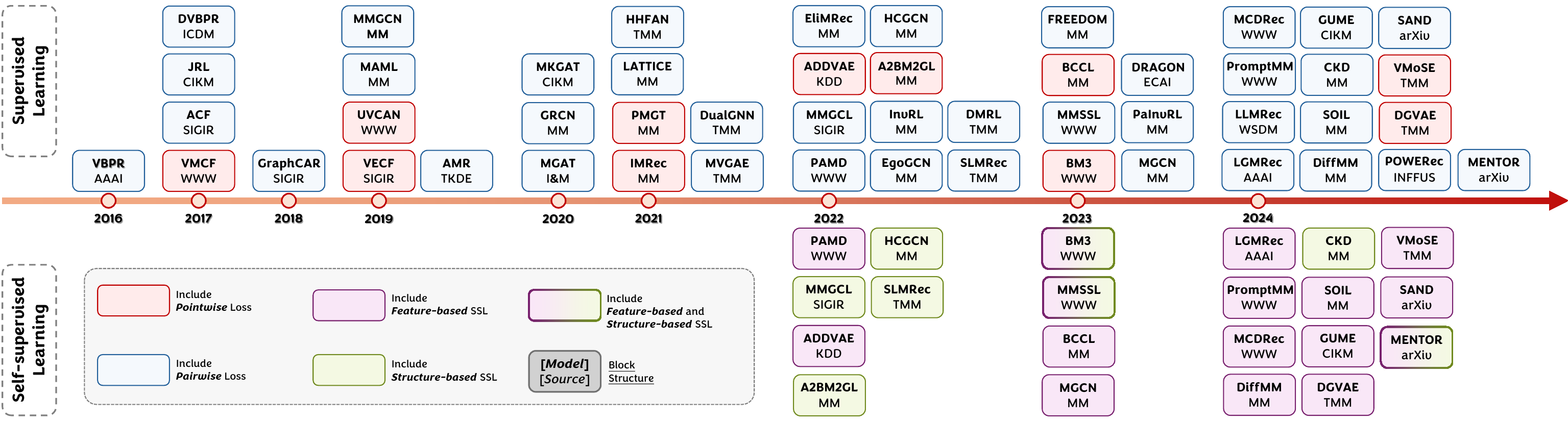}
    \caption{Taxonomy of Loss Functions.}
    \label{fig:Summary of Loss Functions}
\end{figure*}

\subsection{Taxonomy}
To provide researchers with a clearer framework for selecting the most effective loss function for their work, we have systematically categorized recent works, as Figure~\ref{fig:Summary of Loss Functions}. Most existing MRs methods leverage pairwise loss function, only VMCF\cite{park2017also}, VECF\cite{chen2019personalized}, UVCAN\cite{liu2019user}, IMRec\cite{xun2021we}, PMGT\cite{liu2021pre}, ADDVAE\cite{tran2022aligning}, A2BM2GL\cite{cai2022adaptive}, BM3\cite{zhou2023bootstrap}, BCCL\cite{yang2023modal}, DGVAE\cite{zhou2024disentangled}, and VMoSE\cite{yi2024variational} adopt point wise loss function. Furthermore, we provide a specific analysis for self-supervised learning of all recent MRS works.

\textbf{Feature-based SSL:} PAMD \cite{han2022modality} uses self-supervised signals to assist in learning disentangled representation from the feature level. A2BM2GL \cite{cai2022adaptive} proposes aligning disentangled factors learned from ratings and textual content based on regularization and compositional de-attention mechanism. BCCL \cite{yang2023modal} further introduces a bias-constrained data augmentation method to ensure the quality of augmentation samples in contrastive learning. MGCN \cite{yu2023multi}, LGMRec \cite{guo2024lgmrec}, SOIL\cite{su2024soil}, GUME\cite{lin2024gume} use contrastive learning from the feature level to improve the representation quality. PromptMM \cite{wei2024promptmm} introduces a learnable prompt module that dynamically bridges the semantic gap between the multi-modal context encoding in the teacher model and the collaborative relation modeling in the student model. MCDRec \cite{ma2024multimodal}, DiffMM \cite{jiang2024diffmm}, DGVAE \cite{zhou2024disentangled}, and VMoSE \cite{yi2024variational} optimize KL divergence to compels posterior distribution closer to the prior distribution. SAND \cite{he2024boosting} uses self-supervised signals to distinguish modal-unique and modal-generic representations.

\textbf{Structure-based SSL:} MMGCL \cite{yi2022multi} uses modality masking and modality edge dropout to enhance the modal consistency of representations through self-supervision from a structural perspective. A2BM2GL \cite{cai2022adaptive} utilizes the attention mechanism to dynamically learn the importance weights of short-range and long-range neighboring nodes jointly to obtain more expressive representations. HCGCN \cite{mu2022learning} uses contrastive loss from a structural perspective to coordinate multimodal features. SLMRec \cite{tao2022self} uses contrastive learning from a variety of common structural perspectives to improve representation quality.

\textbf{Mixed SSL:} BM3 \cite{zhou2023bootstrap} uses multiple common self-supervised signals at the feature level and structure level to enhance representation. MMSSL \cite{wei2023multi} and MENTOR \cite{xu2024mentor} align different modal representations at the feature level and enhance representations using common contrastive learning from the feature perspective.

% \input{Tab/Loss_Function}

% \red{Todo}

\section{Future Direction}
\label{sec: Future Direction}
In this section, we explore potential future directions for the field of MRS. Our goal is to stimulate and foster further research, development, and innovation within this rapidly evolving domain. By identifying and discussing emerging trends and unresolved challenges, we hope to inspire continued academic inquiry and technological advancement that will drive the next wave of breakthroughs in MRS.

\subsection{Towards Unified MRS Model}
Existing models in the MRS field typically segregate the feature extraction and representation encoding into two distinct processes. The former often leverages existing pre-trained models, while the latter receives more focused attention. However, this segregated process results in inherent multimodal noise, leading to a disconnect between the extracted features and their subsequent encoding. Consequently, there is an urgent need for a unified model that integrates these processes more cohesively. Such a unified model will enhance the relevance and efficiency of the multimodal data, thereby improving the overall accuracy of the recommendation systems in leveraging complex multimodal information.

\subsection{Resolves Cold-start Problem}
Existing RS models typically operate under the assumption of a fixed number of users and items during the training phase, which poses challenges when adapting to the continuous influx of new data. In practical settings, these models are deployed within dynamic environments where new user-item interactions, as well as new users and items, are frequently introduced—a phenomenon often referred to as the cold-start problem. To address this issue, it is beneficial to leverage multimodal information in cold-start scenarios. The integration of diverse data modalities—such as textual descriptions, images, and metadata associated with users and items—can significantly enhance the model's ability to understand and predict the preferences of new users and the attributes of new items effectively. By exploiting multimodal information, RS models can generate more accurate and reliable recommendations even when confronted with limited interaction data, thereby improving their adaptability and performance in dynamically evolving environments.

\subsection{Towards Richer Variety of Modalities}
Existing MRS models demonstrate considerable effectiveness in utilizing textual and visual modalities. However, the burgeoning richness of information available on the Internet presents opportunities for leveraging a broader variety of modalities. This expansion includes auditory, olfactory, and kinesthetic data, among others, which can enrich the understanding and personalization capabilities of RS models.

In response to this evolving landscape, there is an urgent need for MRS models that can effectively integrate and synthesize information from these varied data sources. By harnessing a wider array of modal information, MRS models can achieve a more holistic understanding of user preferences and item characteristics. This comprehensive approach not only enhances the accuracy and relevance of recommendations but also significantly improves user engagement and satisfaction by catering to diverse sensory preferences and interaction styles. Thus, developing MRS models capable of effectively processing and integrating multiple modalities is crucial for advancing the state of the art in RS.

% \red{Todo}
\section{Conclusion}
The primary goal of this survey is to thoroughly examine the recent advancements in MRS and provide a technical analysis of various models. Our discussion categorizes existing MRS models into four critical aspects: Feature Extraction, Encoder, Multimodal Fusion, and Loss Function. Moreover, we review the technical contributions of existing works and explore potential future avenues for advancing and refining MRS technologies. Our contributions extend beyond mere summarization. We tailored technological taxonomy and proposed potential directions for future research. This survey serves as a valuable resource for researchers in the field, offering insights and guidance into the evolving landscape of multimedia recommendations.

\bibliographystyle{IEEEtran}

\end{document}